\documentclass[
superscriptaddress,
amsmath,amssymb,
aps,
pre,10pt
]{revtex4-2}

\usepackage{graphicx}
\usepackage{dcolumn}
\usepackage{bm}
\usepackage{placeins}
\usepackage[dvipsnames]{xcolor}

\usepackage{soul}

\begin{document}

\title[Article Title]{
Impact of ageostrophic dynamics on the predictability of Lagrangian trajectories in surface-ocean turbulence
}

\author{Michael Maalouly}
 \thanks{Corresponding author}\email{michael.maalouly@univ-lille.fr}
\affiliation{Univ. Lille, ULR 7512 - Unit\'e de M\'ecanique de Lille Joseph Boussinesq (UML), F-59000 Lille, France}

\author{Guillaume Lapeyre}
  \affiliation{LMD/IPSL, CNRS, Ecole Normale Sup\'erieure, PSL Research University, 75005 Paris, France}

 \author{Stefano Berti}
  \affiliation{Univ. Lille, ULR 7512 - Unit\'e de M\'ecanique de Lille Joseph Boussinesq (UML), F-59000 Lille, France}

\begin{abstract}
{

Turbulent flows at the surface of the ocean deviate from geostrophic equilibrium on scales smaller than about $10$~km. These scales are associated with important vertical transport of active and passive tracers, and should play a prominent role in the heat transport at climate scales and for plankton dynamics. 
Measuring velocity fields on such small scales is notoriously difficult but new, high-resolution satellite altimetry is starting to reveal them. 
However, the satellite-derived velocities essentially represent the geostrophic flow component, and the impact of unresolved ageostrophic motions on particle dispersion needs to be understood to properly characterize transport properties.
Here, we investigate ocean fine-scale turbulence using a model that represents some of the processes due to ageostrophic dynamics.
We take a Lagrangian approach and focus on the predictability of the particle dynamics, comparing trajectories advected by either the full flow or by its geostrophic component only.
Our results indicate that, over long times, relative dispersion is marginally affected 
after filtering out the ageostrophic component.
Nevertheless, advection by the 
geostrophic-only flow leads to an overestimation of the typical pair-separation rate, and to a bias on trajectories (in terms of displacement from the actual ones), whose importance grows with the Rossby number. 
We further explore the intensity of the transient particle clustering induced by ageostrophic motions and find that it can be significant, even for small flow compressibility.
Indeed, we show that clustering is here due to the interplay between compressibility and persistent flow structures that trap particles, enhancing their aggregation.

}

\end{abstract}

\maketitle

\section{Introduction}\label{sec:intro}

Ocean dynamics involve an extremely wide range of scales, from planetary scales, where energy is injected through solar and tidal forcing, to millimeter ones, where energy is dissipated by viscosity~\cite{Vallis2017}. At scales larger than about $100$~km, oceanic flows are quasi two-dimensional (2D), due to the importance of Earth's rotation and ocean stratification. They nearly satisfy geostrophic balance, i.e. the balance between Coriolis and pressure forces in the momentum equation of motion.
At small scales, instead, they considerably deviate from this equilibrium to become fully three-dimensional (3D), isotropic and turbulent. The main nondimensional control parameter, in this context, is the Rossby number $Ro=U/(fL)$, where $U$ and $L$ are typical horizontal velocity and length scales, respectively, and $f$ is Coriolis frequency (see, e.g., \cite{Vallis2017}); at large scales $Ro \ll 1$, while at small ones $Ro \gg 1$. 
Scales of about $O(100)$~km form the mesoscale range, corresponding to ocean eddies that evolve on timescales of a few weeks to months. They represent the largest fraction of the ocean kinetic energy and can account for intense horizontal transport of heat and concentrations of biogeochemical tracers ~\cite{zhang2014oceanic}. The associated Rossby number is small, and vertical velocities are relatively weak~\cite{klein2009oceanic,ZQKT2019}.
Smaller scales, between $O(1)$ and $O(10)$~km and with a temporal variation of about $1$~day, have larger Rossby numbers, up to order $1$, and are referred to as submesoscales~\cite{McWilliams2016}, or sometimes also as fine scales~\cite{Morrow_etal_2019,barcelo2021fine}. 
In terms of flow structures, they appear as  fronts and filaments, surrounding eddies, and are much more confined at the surface with respect to mesoscales. According to several theoretical and numerical studies and {\it in situ} campaigns ~\cite{McWilliams2016,klein2009oceanic}, their related vertical velocities are an order of magnitude larger than those of mesoscales. Hence, through their role on vertical transport, submesoscales should be essential to climate and biogeochemical processes in the ocean~\cite{klein2009oceanic,su18}.

Measuring velocity fields at submesoscales  represents a challenge, due to their small size and fast evolution, and their direct observation on planetary scales is still lacking. Most of the available information on the dynamics of such flows comes from surface drifters. This includes remarkable features such as: enhanced dispersion rates~\cite{LE2010,BDLV2011,Poje_etal_2014,CLPSZ2017} pointing to energetic submesoscales; 
Lagrangian-tracer clustering~\cite{Dasaro_etal_2018,vic2022} 
pointing to high vorticity and divergence values and strong departure from geostrophy; 
evidence of a direct energy cascade from large to small scales in addition to the more usual inverse cascade~\cite{balwada2022direct}, which may help to shed light on the energy transfers from the forcing to the dissipation scale. 
It is worth noting, however, that bridging the Lagrangian results to the Eulerian framework can be delicate in practical situations, due to limited particle numbers and imperfect flow symmetries (homogeneity and isotropy, as theoretically required). Eulerian measurements, on the other hand, rely on satellite altimetry, which measures sea surface height (SSH). 
From the latter, the velocity field is obtained by applying geostrophic balance. This approach allowed considerable progress on the understanding of mesoscale dynamics~\cite{fu2010eddy,morrow2023ocean}, but the spatial resolution [$O(100)$~km] of conventional instruments has not, so far, allowed access to the submesoscale range. 
A major step forward on this point is expected from the Surface Water and Ocean Topography (SWOT) mission~\cite{Morrow_etal_2019,fu2024surface}. 
This new satellite, launched at the end of 2022, has indeed started measuring 
SSH at a spatial resolution of about $1$~km~\cite{fu2024surface}, which represents two orders of magnitude of improvement.
The innovation brought about by SWOT suggests the possibility to observe the energy cascade over a much broader range of scales, and to achieve a global view of turbulent transport properties at the ocean surface at fine scales. Nevertheless, it also raises important questions in relation to the interpretation of such high-resolution data. In particular, the instrument still measures SSH, and it is unclear down to what length scale geostrophic balance is valid, thus allowing to obtain an accurate velocity field from it~\cite{Rocha2016,Wang2019}. Moreover, geostrophic flows are by definition nondivergent, which prevents the quantification of convergence events and tracer-particle clustering, crucial to the modeling of plankton dynamics and the prediction of pollutant spreading or accumulation. The impact of unresolved ageostrophic motions on transport and dispersion features thus needs to be assessed.

Quasi-geostrophic (QG) theory, derived from an asymptotic expansion of the fundamental governing equations to lowest order in $Ro$~\cite{Vallis2017}, provides a good description of ocean dynamics in the mesoscale range, where rotation and stratification are still quite important and the Rossby number very small. 
Different improvements of this theory have been proposed to account for the observed energetic content of submesoscales close to the surface, particularly by including mixed-layer instabilities~\cite{Callies_etal2016,BL2021}, or by assuming surface dynamics intensified by the action of large-scale strain on buoyancy fronts~\cite{lapeyre2006}. The latter approach gives rise to the surface quasi-geostrophic (SQG) model~\cite{HPGS1995,lapeyre2017},  
which appears appealing considering that, in spite of its simple mathematical formulation, it produces kinetic energy spectra considerably shallower than purely QG ones. 
Other important features of submesoscale flows, however, are still not captured by these improved models, such as the asymmetry of vorticity statistics, with cyclones prevailing over anticyclones~\cite{Rudnick2001,RK2010,Shcherbina_etal_2013,buckingham2016}, and the occurrence of Lagrangian convergence events~\cite{Jacobs_etal_2016,Dasaro_etal_2018,Berta_etal_2020}. These are triggered by ageostrophic motions at fronts. 
A natural strategy to represent their dynamics is to develop the fundamental equations to next order in $Ro$, with respect to the QG approximation. A model developed this way, which allows to reproduce 
a cyclone-anticyclone asymmetry, is the surface semi-geostrophic one~\cite{ragone_badin_2016}. 
Another one, based on SQG dynamics and perhaps better documented, is the SQG$^{+1}$ model~\cite{HSM2002}, which 
we recently considered in an oceanographic setup~\cite{maaloulyetal2023}. In the latter study we have shown that this model can account for both the Eulerian and Lagrangian properties mentioned above. In particular, its dynamics are characterized by a dominance of cyclones over anticyclones, and are responsible of  temporary tracer-particle clustering in cyclonic frontal regions, with the intensity of these phenomena increasing with the Rossby number. However, relative dispersion properties were found to 
depend only marginally on $Ro$.

In this work, we adopt again the SQG$^{+1}$ model to numerically study ocean fine-scale turbulence in the presence of ageostrophic motions, with the aim of investigating the impact of the latter on Lagrangian transport. 
Ageostrophy will then be taken into account as a first-order correction to the geostrophic terms in the model equations. 
Our main goal is to dispose of an idealized modeling framework to explore the accuracy of velocity fields similar to those from satellites like SWOT. To address this point, we
focus on Lagrangian predictability, comparing trajectories of particles advected by either the full flow or its geostrophic part only (i.e. where ageostrophic motions are
artificially removed), 
which should be closer to that measured by satellite altimetry. 
Note that, due to the dynamical couplings between the surface temperature and velocity fields in the system, this is not the same as considering flows at $Ro=0$. 
Specifically, we perform a systematic comparison of the turbulent dispersion properties of the two kinds of trajectories, in terms of two-particle statistics, mainly relying on Lagrangian Lyapunov exponents of different types. 
The results are in agreement with earlier findings, obtained comparing particle advection in 
weakly-ageostrophic-flow simulations at different Rossby numbers~\cite{maaloulyetal2023}, about the weak effect of the ageostrophic velocity on relative dispersion. However, they also highlight that advection by the geostrophic-only flow
tends to overestimate the typical pair-separation rate. Moreover, we show that 
filtering out the ageostrophic flow causes a bias on trajectories, whose importance grows with $Ro$, and we quantify the 
scale-by-scale dispersion rate between the full and geostrophic-only advection models. 
We further provide a characterization of the temporary particle clusters that form due to ageostrophic motions. 
In particular, we find that, while compressibility is 
small in our simulations, 
the intensity of clustering can be substantial. 
Our analysis indicates that, in the SQG$^{+1}$ system, clustering is essentially due to the interplay between the (small) flow compressibility and the existence of long-lived structures that trap particles, increasing their accumulation. 

This article is organized as follows. In Sec.~\ref{sec:model} we introduce the Eulerian flow model and the Lagrangian dynamics, as well as the simulation settings adopted. The numerical results are presented in Sec.~\ref{sec:results}. There, we first characterize the main turbulent features of the full flow and of its 
geostrophic-only counterpart (Sec.~\ref{sec:eulerian}). Then we consider Lagrangian statistics for tracers advected by either the complete velocity field or by its geostrophic component. We start by discussing the impact of 
removing the ageostrophic flow on the relative-dispersion process (Sec.~\ref{sec:lagr_disp}), and we then examine the small-scale particle dynamics in terms of Lyapunov exponents, particularly focusing on the characterization of clustering in the full flow (Sec.~\ref{sec:clustering}). Finally, discussions and conclusions are presented in Sec.~\ref{sec:conclusion}.

\section{Model and numerical simulations}\label{sec:model}

To describe turbulent dynamics in the presence of ageostrophic motions, we adopt the SQG$^{+1}$ model~\cite{HSM2002}.This was first introduced in an atmospheric context to account for the cyclone-anticyclone asymmetry emerging in rotating stratified fluids at finite Rossby numbers, but not reproduced by QG models. 
In its oceanic formulation, it was recently shown to be a good minimal model for reproducing ageostrophic effects in the fine-scale range~\cite{maaloulyetal2023}. 
The model is obtained from an expansion at next order in $Ro$, with respect to QG theory, of the momentum and buoyancy evolution equations, 
within the Boussinesq and hydrostatic approximations (also known as primitive equations). It constitutes an extension of the SQG system~\cite{HPGS1995,lapeyre2017}, assuming that the flow dynamics are entirely driven by the advection of buoyancy at the surface. The relevance of SQG-like dynamics to upper-ocean turbulence is well documented~\cite{lapeyre2006,lapeyre2017} and mainly involves the occurrence of energetic submesoscales, but also the consequent enhancement of the pair-separation rate of Lagrangian particles~\cite{FBPL2017,BL2021}, and the evolution of phytoplankton diversity~\cite{Perruche_etal11}.

\subsection{Dynamical equations}

The main dynamical equation of the SQG$^{+1}$ model  states that surface temperature (or buoyancy) is conserved along the surface flow,
\begin{equation}
\partial_t \theta^{(s)} + \bm{u}^{(s)} \cdot \bm{\nabla} \theta^{(s)} = 0,
\label{eq:sqgp1_temp}
\end{equation}
where $\theta(\bm{x},t)$ is the temperature fluctuation field. Here and in the following we adopt nondimensional units. The vertical coordinate is $-\infty < z \leq 0$, and the superscript $(s)$ indicates quantities evaluated at the surface ($z=0$). 
The total velocity field is given by the sum of the geostrophic component $\bm{u}_g$ (computed at the lowest order in $Ro$) and an ageostrophic one $\bm{u}_{ag}$ (at next order in $Ro$), so that
\begin{equation}
\bm{u}=\bm{u}_g + Ro \, \bm{u}_{ag}. 
\label{eq:sqgp1_veltot}
\end{equation}
The geostrophic velocity is obtained from the streamfunction $\phi$ as $\bm{u}_g=\left( -\partial_y \phi, \partial_x \phi \right)$, where $x$ and $y$ denote the horizontal coordinates.  Setting $Ro=0$ in Eqs.~(\ref{eq:sqgp1_temp}) and~(\ref{eq:sqgp1_veltot}), the SQG model is recovered. 
Here below we recall the main steps leading to 
the expression of $\bm{u}_{ag}$
; more details about the full derivations can be found in~\cite{HSM2002,Weiss2022}.
At first order, the interior potential vorticity vanishes, i.e.
\begin{equation}
    \frac{\partial^2\phi}{\partial x^2}+\frac{\partial^2\phi}{\partial y^2}+\frac{\partial^2\phi}{\partial z^2}=0.
\end{equation}
The geostrophic streamfunction $\phi$ is then obtained from surface temperature, with the boundary conditions $\partial_z \phi|_{z=0}=\theta^{(s)}$ and $\partial_z \phi \rightarrow 0$ for $z \rightarrow -\infty$, i.e.
\begin{equation}
\phi = \mathcal{F}^{-1}\left[ \frac{\mathcal{F}(\theta^{(s)})}{k}e^{kz} \right].
\label{eq:sqgp1_phi_cap}
\end{equation}
In this equation, $\theta$ is taken at lowest order, $\mathcal{F}$ stands for the horizontal Fourier transform and $k$ for the horizontal wavenumber modulus. Note that $\theta=\partial_z\phi=\mathcal{F}^{-1}\left[ \mathcal{F}(\theta^{(s)})e^{kz} \right]$.\\ 
The ageostrophic velocity $\bm{u}_{ag}$ is expressed as the sum of two contributions, $\bm{u}_\varphi$ and $\bm{u}_a$, so that
\begin{equation}
    \bm{u}_{ag} = \bm{u}_\varphi + \bm{u}_a.
    \label{eq:eq:sqgp1_u_ag}
\end{equation}
Remarkably, these SQG$^{+1}$ ageostrophic velocity components, like the geostrophic one, can be computed from the temperature field $\theta$. 
Indeed, they can be written as 
\begin{equation}
\bm{u}_\varphi=\left( -\partial_y \varphi, \partial_x \varphi \right), \; \bm{u}_a=-\partial_z \bm{A},
\label{eq:sqqp1_uphi_ua}
\end{equation}
where $\varphi$ and $\bm{A}$ are related to surface and lower-order quantities by:
\begin{equation}
\varphi = \frac{\theta^2}{2} - \mathcal{F}^{-1}\left\{ \frac{\mathcal{F}\left[\theta^{(s)} 
(\partial_z \theta)^{(s)} \right]}{k}e^{kz} \right\},
\label{eq:sqgp1_phi}
\end{equation}
\begin{equation}
\bm{A} = -\theta \bm{u}_g + \mathcal{F}^{-1}\left[ \mathcal{F}(\theta^{(s)} \bm{u}_g^{(s)}) e^{kz} \right],
\label{eq:sqgp1_A}
\end{equation}
again with $\theta$ taken at lowest order. 
Functions $\varphi$ and $\bm{A}$ are such that $\partial_z\varphi=0$ and $\bm{A}=\bm{0}$ at $z=0$.
Note that $\bm{u}_a$ has both a rotational and a divergent component from (\ref{eq:sqgp1_A}), while $\bm{u}_{\varphi}$ is nondivergent.

The idealized character and relatively simple mathematical formulation of this model represent a strong advantage. One of its limitations, however, is that other types of ageostrophic dynamics, further deviating from geostrophic equilibrium, cannot be taken into account. Among these, high-frequency motions (internal gravity waves and tides), in particular, may be expected to also play a relevant role on submesoscale turbulence~\cite{yu2019surface,wang2022deep,arbic2022near}. Addressing the impact of such processes on Lagrangian dynamics is an interesting point, but it is left for future work, as it would require more realistic simulations.

In this work we are interested in the trajectories of Lagrangian particles at the surface, for advection realized by either the full flow $\bm{u}$ (obtained from integration of the above equations, and evaluated at $z=0$) or its geostrophic component $\bm{u}_g$, once the ageostrophic velocity $\bm{u}_{ag}$ is, {\it a posteriori}, filtered out from $\bm{u}$.
The particle equations of motion then are, respectively, 
\begin{equation}
    \frac{d\bm{x}}{dt} = \bm{u}(\bm{x}(t),t),
    \label{eq:motiontracers_f}
\end{equation}
\begin{equation}
    \frac{d\bm{x}_g}{dt}=\bm{u}_g(\bm{x}_g(t),t),
    \label{eq:motiontracers_g}
\end{equation}
where $\bm{u}=(u,v)$ (and similarly for $\bm{u}_g$), and $\bm{x}(t)$ and $\bm{x}_g(t)$ denote the horizontal position of a particle evolving in either of the two flows. In these equations, velocities come from the same simulation, and are evaluated at the same time $t$. In the following, to ease the distinction between results obtained from Eq.~(\ref{eq:motiontracers_f}) or Eq.~(\ref{eq:motiontracers_g}), we will also use the subscript $f$ to indicate quantities computed using the full flow ($\bm{u}_f \equiv \bm{u}$).

\subsection{Numerical settings}

To obtain the full Eulerian velocity field, we numerically integrate Eq.~(\ref{eq:sqgp1_temp}), with Eqs.~(\ref{eq:sqgp1_veltot}-\ref{eq:sqgp1_A}), 
on a doubly periodic square domain of side $L_0=2\pi$ at resolution $N^2=1024^2$, by means of a pseudospectral method~\cite{maaloulyetal2023}, for increasing Rossby numbers (starting from $Ro=0$).
The initial condition corresponds to a streamfunction, in Fourier space, with random-phase and small-amplitude modes. 
In order to reach a statistically steady state, we consider the forced and dissipated version of Eq.~(\ref{eq:sqgp1_temp}).  
Specifically, we add on the right-hand side of the equation a random ($\delta$-correlated in time) forcing acting 
over a narrow range of wavenumbers $4\leq k_f \leq 6$ (and whose intensity is $F=0.02$), as well as a hypofriction term $-\alpha \bm{\nabla}_H^{-2} \theta$ to remove energy from the largest scales, and a hyperdiffusion term $-\nu\bm{\nabla}_H^4\theta$ to assure small-scale dissipation and numerical stability. 
For the dissipative terms we set  $\alpha=0.5$ and we determine $\nu$ according to the condition $k_{max} l_\nu \gtrsim 6$, with $l_\nu$ the dissipative scale (estimated for $Ro=0$). While such values result in quite large dissipation terms, which reduce the number of active scales, they were found to be needed to control numerical stability at the largest $Ro$ values. The compressibility of the SQG$^{+1}$ horizontal flow, in fact, produces intense gradients that are difficult to resolve. The largest Rossby number we could reach is $Ro=0.075$ 
(note that this is just an {\it a priori} value, see next section for an estimation of $Ro$ based on the dynamics). 
A third-order Adams-Bashforth scheme  is used to advance in time Eq.~(\ref{eq:sqgp1_temp}), with forcing and dissipation terms. 
The time step was set to the quite small value $dt=10^{-4}$, 
ensuring temporally converged results for all the Rossby numbers explored.
   
In all the SQG$^{+1}$ simulations we compute Lagrangian trajectories according to both Eq.~(\ref{eq:motiontracers_f}) and Eq.~(\ref{eq:motiontracers_g}), where the ageostrophic flow component is excluded. Clearly, for the SQG case ($Ro=0$), the velocity field is purely geostrophic. 
The Lagrangian evolution equations are integrated using a third-order Adams-Bashforth scheme and bicubic interpolation in space of the velocity field at particle positions. An infinite domain is assumed, 
the Lagrangian velocities outside the computational box being computed using the spatial periodicity of the Eulerian flow. 
We consider $N_p=49152$ particles, whose initial positions correspond to a regular arrangement of $M=128 \times 128$ triplets over the entire domain. 
Each triplet forms an isosceles right triangle, with a particle pair along $x$ and one along $y$, 
both of which are characterized by an initial separation $R(0)=\Delta x/2$ (with $\Delta x$ the grid spacing). 
Particles are injected into the considered flow once this has reached statistically stationary conditions. 
Dispersion statistics are computed relying only on original pairs (which are $32768$ in each simulation). We checked that 
pair separation statistics are independent of the initial orientation (along $x$ or $y$ direction) of the pairs, and that the results are robust with respect to the number of pairs used.

\section{Results}\label{sec:results}

\subsection{Eulerian properties of the turbulent flow and its geostrophic component}
\label{sec:eulerian}

\begin{figure*}[htbp]
\centering
\includegraphics[width=1\textwidth]{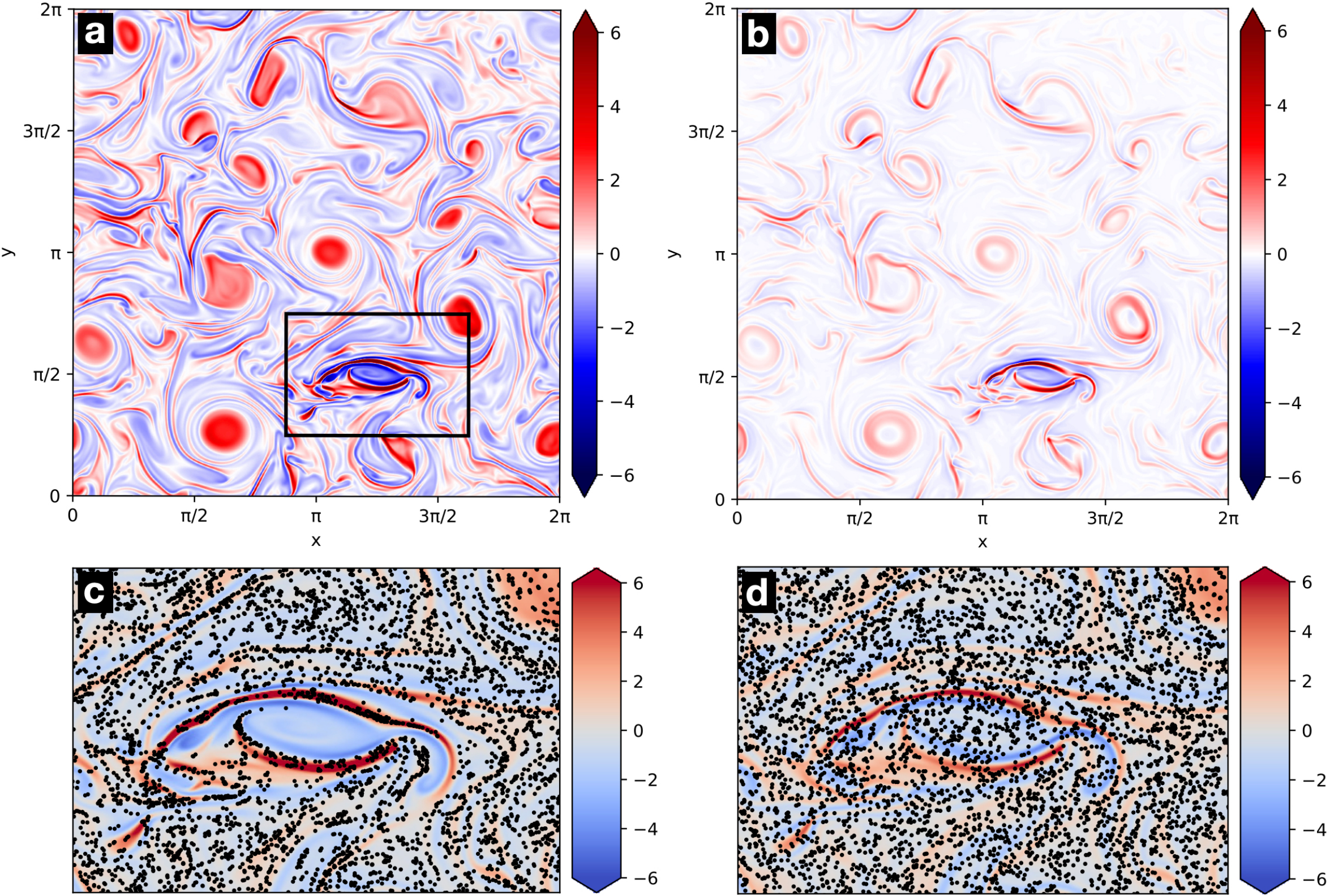}
\caption{(a) Vorticity field $\zeta_{f}$ for the SQG$^{+1}$ system for $Ro=0.0625$ at a specific time $t_*$ during the statistically stationary state. (b) Difference field $\Delta \zeta = \zeta_{f} - \zeta_{g}$, where $\zeta_g$ is the geostrophic component of vorticity. 
Panels (c) and (d) show the distribution of particles at time $t_*$ in the region corresponding to the black rectangle in (a), for advection realized by either the full flow (c), or its geostrophic component (d). 
In (c) and (d) the full and geostrophic vorticity fields, respectively, are shown in color.
In all panels, vorticity is normalized by the rms value of $\zeta_{f}$. 
}
\label{fig:vorticity_diff}
\end{figure*}

For nonzero Rossby number, the SQG$^{+1}$ flow is characterized by well defined, mainly cyclonic, eddies of different sizes, and sharp gradients along filament-like structures. This is illustrated in Fig.~\ref{fig:vorticity_diff}a, which shows the (full) vorticity field $\zeta_f=\partial_x v - \partial_y u$, normalized by its root-mean-square (rms) value $\zeta_f^{\mathrm{rms}}$, for $Ro=0.0625$ at an instant of time $t_*$ in the statistically steady state reached by the system after a transient. 
The rms vorticity is found to be close to $\zeta_f^{\mathrm{rms}} \approx 10$, and to slightly grow with $Ro$ but overall to weakly depend on it. It provides an {\it a posteriori} measure of the Rossby number as 
$Ro \, \zeta_f^{\mathrm{rms}}
\lesssim 1$, 
suggesting that the small scales of our flows can be interpreted as submesoscales.

The presence of strong gradients in the horizontal flow (as visualized by $\zeta_f$) is a generic feature due to the ageostrophic velocity components~\cite{maaloulyetal2023}. This can be deduced from  Fig.~\ref{fig:vorticity_diff}b where  we show the difference field $\Delta \zeta = \zeta_{f} - \zeta_{g}$ (i.e. the ageostrophic vorticity), again normalized by $\zeta_f^{\mathrm{rms}}$. Positive values of $\Delta \zeta$ can be seen at the periphery of cyclonic eddies and along extended filaments. This implies that  the full cyclonic vorticity, $\zeta_f$, is stronger 
than its geostrophic component, $\zeta_g$.

Filtering out the ageostrophic velocities has consequences also on Lagrangian dynamics. 
For instance, when initially uniformly distributed tracer particles are advected by either the full or the geostrophic-only flow, important qualitative differences emerge, such as the occurrence of clustering when ageostrophic fluid motions are included (Fig.~\ref{fig:vorticity_diff}c). 
For the geostrophic flow, instead, no sign of clustering is observed 
(Fig.~\ref{fig:vorticity_diff}d), as expected due to the nondivergent character of this flow. 
We will discuss particle dispersion properties and clustering in Sec.~\protect{\ref{sec:lagr_disp}} and Sec.~\protect{\ref{sec:clustering}}.

\begin{figure}[htbp]
\centering
\includegraphics[width=0.6\textwidth]{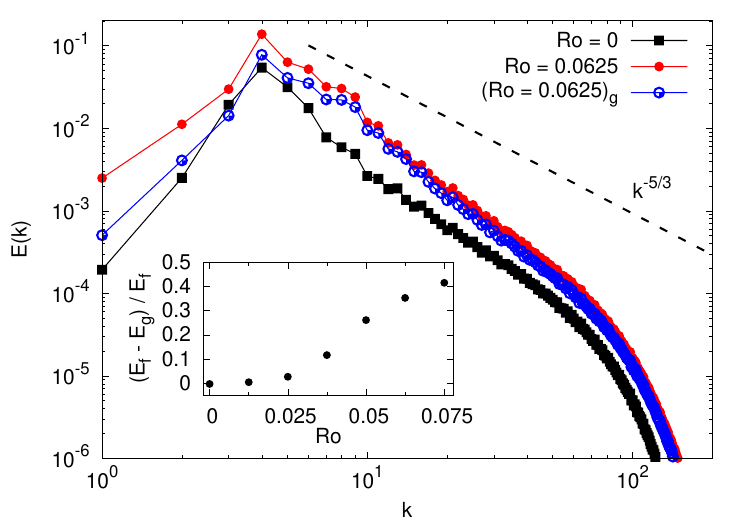}
\caption{Kinetic energy spectra, temporally averaged over several flow realizations in the statistically steady state for $Ro = 0$, $Ro = 0.0625$ and $(Ro = 0.0625)_g$ (i.e. 
the kinetic energy of the geostrophic component 
in the simulation at $Ro = 0.0625$). 
The dashed black line corresponds to
$k^{-{5/3}}$, the expected spectrum for SQG turbulence. 
Inset: absolute value of the relative difference of kinetic energy 
computed from the full flow and from its
geostrophic component,
as a function of $Ro$.}
\label{fig:energy_spectra}
\end{figure}

We now examine the statistical features of the Eulerian flows from a more quantitative point of view. 
Figure~\ref{fig:energy_spectra} shows the kinetic energy spectrum $E(k)$, with $k$ the horizontal wavenumber modulus, for three cases: the purely SQG ($Ro=0$) flow, the full SQG$^{+1}$ flow at $Ro=0.0625$ and its geostrophic component [i.e. filtering out $\bm{u}_{ag}$ in Eq.~(\ref{eq:sqgp1_veltot})]. 
In all cases we find that spectra follow power laws $E(k) \sim k^{-\beta}$ over about a decade. 
In an oceanographic context, this means that our simulations resolve both the mesoscale range [$O(100)$~km], here corresponding to spatial scales $\ell \approx 1/k_f$, and the submesoscale range down to length scales of $O(10)$~km, here corresponding to $\ell \approx 1/(10 \, k_f)$. 
For both the $Ro=0$ and full $Ro=0.0625$ cases, the exponent $\beta$ is larger than $5/3$, the value expected for SQG turbulence forced at large scales~\cite{lapeyre2017}.
This fact is found to be general and independent of the Rossby number, with spectral exponents in the range $2.2 \lesssim \beta \lesssim 2.7$ (not shown). 
Its causes are the presence of large persistent structures (of size comparable with the forcing lengthscale), which are known to steepen the spectrum~\cite{HSM2002,CCMV2004,maaloulyetal2023}, but also the important values of the small-scale dissipation coefficients used~\cite{FBPL2017}. 
The kinetic energy of the geostrophic-only flow at $Ro=0.0625$ is found to be lower at all scales than that of the corresponding full flow (Fig.~\ref{fig:energy_spectra}), and the same is true for all the Rossby numbers considered (not shown).
However, it bears strong similarities with the spectrum of the full flow (at the same Rossby number), particularly at scales $k>10$. 
Note, too, that it is quite different from that  of the $Ro=0$ flow.
This provides a first evidence of the fact that, even after filtering out $\bm{u}_{ag}$, traces of the influence of the ageostrophic velocities are still discernible in the geostrophic flow component. In other words, the properties of a genuine, dynamically constrained geostrophic flow are not fully recovered once ageostrophic 
dynamics are (\textit{a posteriori}) removed from the complete flow. 

\begin{figure}[htbp]
\centering
\includegraphics[width=0.6\textwidth]{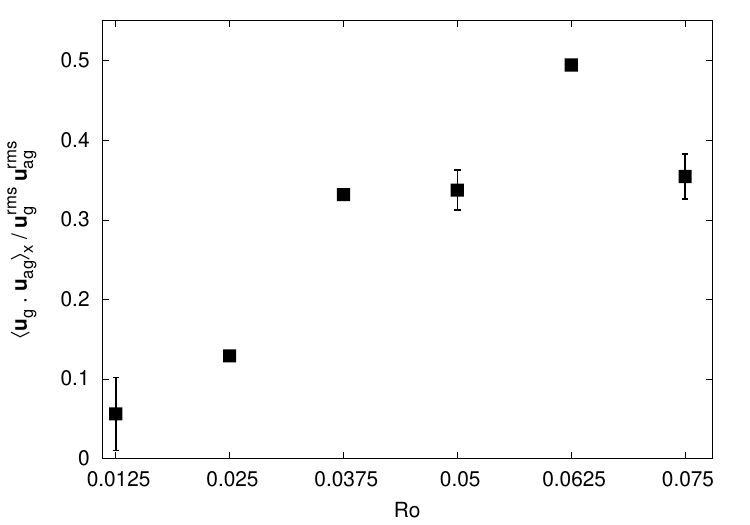}
\caption{
Normalized correlation between the geostrophic and ageostrophic flow components as a function of $Ro$ (where $u_{g}^{\mathrm{rms}}=\sqrt{\langle|\bm{u}_g|^2\rangle}$ and similarly for $u_{ag}^{\mathrm{rms}}$).  
Here, the average is over several flow realizations in statistically steady conditions and error bars are computed as the
difference between the average over the full dataset and over half the dataset. 
}
\label{fig:urms}
\end{figure}

The relative difference between the kinetic energy of the full flow and the 
geostrophic one
$(E_f - E_g)/E_f$ grows with increasing $Ro$ and can reach about $40\%$ at the highest Rossby number (inset of Fig.~\ref{fig:energy_spectra}). Note that these values do not appreciably change when the contributions from the smallest wavenumbers are excluded from the computation of $E_f$ and $E_g$. 
This difference is clearly due to the ageostrophic kinetic energy $E_{ag}=Ro^2\langle |\bm{u}_{ag}|^2/2 \rangle_x$ (with $\langle ... \rangle_x$ a spatial average), but also to the positive correlation between the geostrophic and ageostrophic components of the flow. 
Indeed, the total velocity is $\bm{u}_f = \bm{u}_g + Ro \, \bm{u}_{ag}$, so that $E_f=E_g+E_{ag} + Ro \langle \bm{u}_g \cdot \bm{u}_{ag} \rangle_x$. 
In our simulations, the last term is found to be always positive (Fig.~\ref{fig:urms}), meaning that it contributes to the increase of $E_f$ with respect to $E_g$. As it is proportional to $Ro$, it is also typically larger than $E_{ag}$, due the $Ro^2$ dependence of the latter.

\begin{figure}[htbp]
\centering
\includegraphics[width=0.6\textwidth]{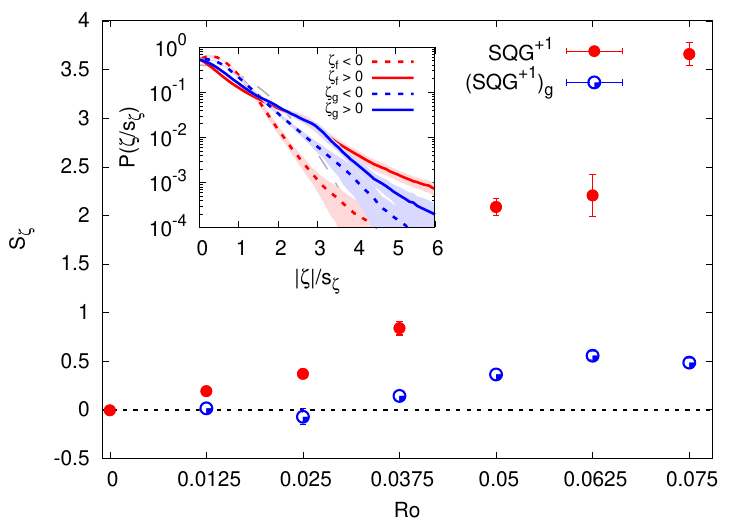}
\caption{
Skewness of vorticity ($S_\zeta=\langle\zeta^3\rangle/\langle\zeta^2\rangle^{3/2}$) as a function of the Rossby number 
averaged over several flow realizations in statistically steady conditions. 
Filled red points correspond to the full SQG$^{+1}$ flow and empty blue points to its geostrophic-only component (SQG$^{+1}$)$_g$.
The error is computed as the difference between the average over the full dataset and over half the dataset. Inset: Probability density function of vorticity $\zeta$ (rescaled by its rms value $s_\zeta$), temporally averaged over several flow realizations in the statistically steady state for $Ro=0.0625$. Here the red and blue colors correspond to the SQG$^{+1}$ and (SQG$^{+1}$)$_g$ cases, respectively. The dashed and solid lines are for $\zeta<0$ and $\zeta>0$, respectively. The shaded areas correspond to the standard deviation of the temporal statistics. The long-dashed gray line represents a Gaussian distribution, for comparison. 
}
\label{fig:vorticity_skewness}
\end{figure}

A distinctive feature of the SQG$^{+1}$ model, absent in the QG and SQG systems, is 
an asymmetry in vorticity statistics, with cyclones prevailing over anticyclones~\cite{HSM2002,maaloulyetal2023}. 
To further investigate the imprints 
of ageostrophic motions,
we consider the probability density function (pdf) of vorticity. 
Figure~\ref{fig:vorticity_skewness} shows vorticity skewness ($S_\zeta$) as a function of $Ro$, for the total flow and its geostrophic component (i.e. filtering out ageostrophic vorticity related to $\bm{u_{ag}}$).  
The corresponding  pdfs $P(\zeta)$ are reported in the inset of Fig.~\ref{fig:vorticity_skewness} (with $\zeta$ rescaled by its rms value $s_\zeta$) for $Ro=0.0625$.  
For the full SQG$^{+1}$ flows, positive
skewness $S_\zeta$, indicative of the predominance of cyclonic structures, characterizes
the vorticity pdf and this effect becomes more important with increasing $Ro$.
The skewness of the vorticity obtained after filtering out $\bm{u}_{ag}$ significantly drops to values that are much closer to zero.
However, it definitely stays positive at large enough Rossby numbers (see also the inset of Fig.~\ref{fig:vorticity_skewness}).
This means that the cyclone-anticyclone asymmetry, though strongly reduced, still persists in the 
geostrophic-only velocity field and highlights, once more, that the latter is different from a purely SQG flow at $Ro=0$.

This result is confirmed by the vorticity pdfs presented in the inset of Fig.~\ref{fig:vorticity_skewness}. The 
right tail ($\zeta_f>0$) is much higher than 
the left one ($\zeta_f<0$). When taking only the geostrophic flow component, there is a decrease of the right tail (and rise of the left one) of $P(\zeta_g)$ 
with respect to the corresponding one in $P(\zeta_f)$.
By looking at the vorticity difference field in Fig.~\ref{fig:vorticity_diff}b, it is possible to see that $\Delta \zeta=\zeta_f-\zeta_g$ is predominantly positive and that a relevant part of the vorticity variation occurs along filamentary structures. In particular, comparison with Fig.~\ref{fig:vorticity_diff}a shows that the intensity of cyclonic ($\zeta>0$) filaments gets lowered 
when $\bm{u}_{ag}$ is filtered out, in qualitative agreement with the behavior {of $P(\zeta)$.}
Such structures play a central role for particle clustering. 
Indeed, drifter studies~\cite{Dasaro_etal_2018} and realistic simulations~\cite{Balwada_etal_2021,vic2022} of submesoscale ocean turbulence indicate that flow convergence (and intense vertical velocities) should take place along cyclonic frontal regions. As we discussed in detail in a previous work~\cite{maaloulyetal2023}, the SQG$^{+1}$ system can be seen as a minimal model capable of accounting for this feature, and giving rise to particle clustering.  
When we compare the particle distributions in Fig.~\ref{fig:vorticity_diff}c and Fig.~\ref{fig:vorticity_diff}d, obtained from advection by the full and 
geostrophic-only flow, respectively, it becomes apparent that substantial variations in the vorticity field reflect in very different particle behaviors. For instance, in the region defined by $\pi \lesssim x \lesssim 3\pi/2$ and $y \approx \pi/2$, we see that particles cluster over an intense positive vorticity filament in the full flow, while this effect completely disappears in the vorticity-weakened, 
geostrophic flow.

\subsection{Lagrangian dispersion}
\label{sec:lagr_disp}

In this section, we compare the particle transport and dispersion properties of the SQG$^{+1}$ flows and of the corresponding (SQG$^{+1}$)$_g$ ones.
Recall that by (SQG$^{+1}$)$_g$ we mean that only the geostrophic component of the flow is used to advect the Lagrangian tracers. 
The analysis presented below relies on both time- and scale-dependent metrics. 

We focus on two-particle statistics, which depend on velocity-field spatial increments and allow to characterize the tracer pair-separation process.  
The most natural way to proceed is perhaps to measure the mean-square relative displacement between two particles [{labeled by $i$ and $j$, and} originally at a given distance $|\bm{x}_i(t_0)-\bm{x}_j(t_0)|=R_0$] as a function of time, i.e. relative dispersion:
\begin{equation}
\langle R^2(t) \rangle = \left\langle |\bm{x}_i(t)-\bm{x}_j(t)|^2 \right\rangle,
\label{eq:rel_disp}
\end{equation}
where $\langle ... \rangle$ is an average over particle pairs.
At sufficiently short times, one expects a ballistic behavior of the form $\langle R^2(t) \rangle \simeq R_0^2 (1+Zt^2)$~\cite{bourgoin_etal_2006,FBPL2017}, where 
$Z=\langle \zeta^2/2\rangle_x$ is enstrophy. 
At very long times, instead, particles typically are at distances much larger than the largest eddies, and a diffusive scaling is expected, $\langle R^2(t) \rangle \sim t$, due to particles experiencing essentially uncorrelated velocities~\cite{LaCasce2008}. 
At intermediate times, when pair separations lie in the inertial range of the flow, relative dispersion should grow exponentially or as a power law, if the kinetic energy spectrum scales as $k^{-\beta}$ with
$\beta>3$ or $\beta<3$, respectively. The first case is generally referred to as a nonlocal dispersion regime, and $\langle R^2(t) \rangle \sim \exp{(2 \lambda_L t)}$, with $\lambda_L$ the maximum Lagrangian Lyapunov exponent. 
In the second case, dispersion is said to be in a local regime, and $\langle R^2(t) \rangle \sim t^{4/(3-\beta)}$~\cite{LaCasce2008,FBPL2017}. 

Another two-particle, fixed-time indicator that can be used to identify dispersion regimes is the kurtosis of the relative distance between particles in a pair~\cite{LaCasce2008,FBPL2017}: 
\begin{equation}
ku(t) = \frac{\langle R^4(t) \rangle}{\langle R^2(t) \rangle^2} .
\label{eq:kurtosis_separation}
\end{equation}
When dispersion is nonlocal (i.e., dominated by the largest  flow structures), rapid (exponential) growth of $ku(t)$ is expected. For local dispersion (meaning controlled by flow features of size comparable with the distance between a pair of particles), the kurtosis should be constant; in particular $ku(t)=5.6$ for Richardson dispersion (the behavior expected for $\beta=5/3$). At larger times, in the diffusive regime, the kurtosis reaches a constant value equal to $2$. 

\begin{figure}[htbp]
\centering
\includegraphics[width=0.49\textwidth]{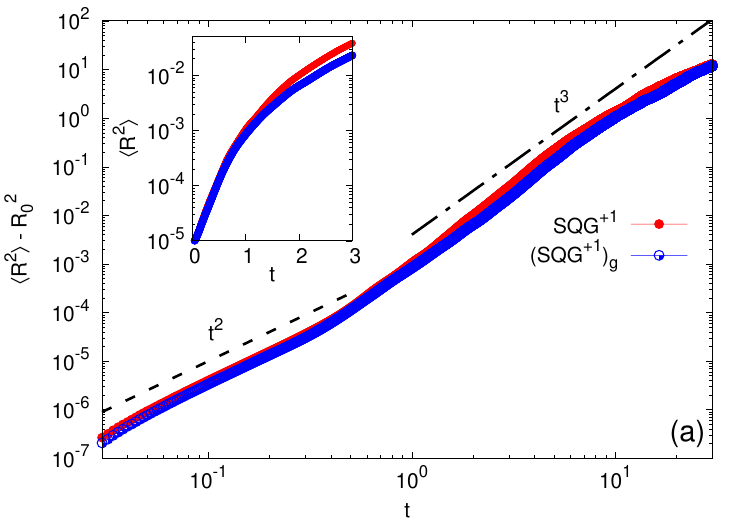}
\includegraphics[width=0.49\textwidth]{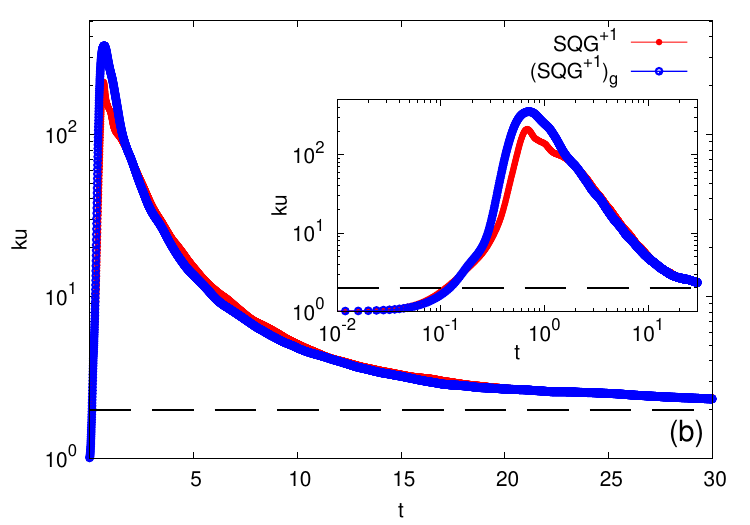}
\caption{(a) Relative dispersion (after subtraction of the initial value) $\langle R^2(t) \rangle - R_0^2$ as a function of time. The $t^3$ (Richardson dispersion) scaling law is the expectation for a kinetic energy spectrum $E(k) \sim k^{-5/3}$, the $t^2$ one is the short-time ballistic expectation. 
Inset: early growth of relative dispersion $\langle R^2(t) \rangle$ versus time in semilogarithmic scale.
(b) Kurtosis of separation as a function of time, {in semilogarithmic scale}. The horizontal dashed line is the expectation $ku=2$ in the diffusive regime. {The inset shows the same in logarithmic scales.}
In both (a) and (b) $Ro=0.0625$; 
colors and symbols have the same meaning as in Fig.~\ref{fig:vorticity_skewness}
}
\label{fig:reldisp}
\end{figure}

Comparison of two-particle statistics as a function of $Ro$ is documented elsewhere~\cite{maaloulyetal2023} and shows that different Rossby numbers lead to qualitatively similar dispersion regimes. Here, at each given Rossby number, we find that two-particle statistics are affected to a limited extent by ageostrophic motions (see Fig.~\ref{fig:reldisp}, for $Ro=0.0625$). 
Indeed, the curves of $\langle R^2(t) \rangle$ 
for particles advected by the full and geostrophic-only flows (Fig.~\ref{fig:reldisp}a) are close, and the same holds for all the values of $Ro$ considered (not shown). 
In both the SQG$^{+1}$ and the (SQG$^{+1}$)$_g$ cases, at short times relative dispersion agrees with the $t^2$ ballistic prediction, the prefactor being close to the enstrophy of the corresponding flow. 
At later times, $\langle R^2(t) \rangle$ is slightly larger in the full flow, 
but the two curves reach the diffusive regime with almost identical values; the same trend is observed at all Rossby numbers, but its importance decreases with $Ro$, and it is hardly detectable for $Ro<0.05$.
At this level, while the effect is small, one may speculate that this slowing down of $\langle R^2(t) \rangle$ in the full-flow case is due to particle trapping in flow convergence regions. 
At intermediate times, relative dispersion grows faster than $t^3$, which is consistent with the spectra of the two flows being steeper than $k^{-5/3}$, but overall the data do not allow to draw quantitative conclusions about the agreement with the predictions for different dispersion regimes. 

The behavior of the kurtosis (Fig.~\ref{fig:reldisp}b) reveals two points. On one side, for both full and 
geostrophic-only flows, the rapid initial growth (up to values $\approx350$) points to nonlocal dispersion. Indeed, for a local dispersion regime, one would instead obtain a stabilization around a constant, much smaller value.
As extensively discussed for SQG$^{+1}$ flows at varying Rossby numbers in a previous work~\cite{maaloulyetal2023}, this is due to the presence of large-scale coherent flow structures that dominate the particle spreading process. On the other side, we find that, except perhaps at the very shortest times, $ku(t)$ grows more rapidly and to higher values in the geostrophic-only flow. While the difference is small, it is clearly detectable, and it is observed also at other Rossby numbers (not shown). This implies that the dispersion regime is more strongly nonlocal 
when particles are advected by the geostrophic component of the flow only
(a result that is difficult to infer from relative dispersion alone).

Fixed-scale indicators are often preferred to fixed-time ones, in order to identify dispersion regimes~\cite{CV2013}. 
For this reason, we now examine the finite-size Lyapunov exponent (FSLE)~\cite{ABCCV1997,CV2013}, which is a scale-by-scale dispersion rate, and is defined as
\begin{equation}
\lambda(\delta)=\frac{\ln~r}{\langle \tau(\delta) \rangle},
\label{eq:FSLE}
\end{equation}
where the average is over all pairs and $\tau (\delta)$ is the time needed for the separation to grow from $\delta$ to a scale $r\delta$ (with $r>1$). 
Dimensionally it is possible to relate the FSLE to the exponent $\beta$ of the kinetic energy spectrum. 
For $\beta>3$ (i.e. in the nonlocal dispersion regime), the FSLE should be constant, $\lambda(\delta)=\lambda_L$. 
When dispersion is local ($\beta<3$), it should have a power-law dependence $\lambda \sim \delta^{(\beta-3)/2}$, while in the diffusive regime one expects $\lambda (\delta) \sim \delta^{-2}$.

\begin{figure}[htbp]
\centering
\includegraphics[width=0.6\textwidth]{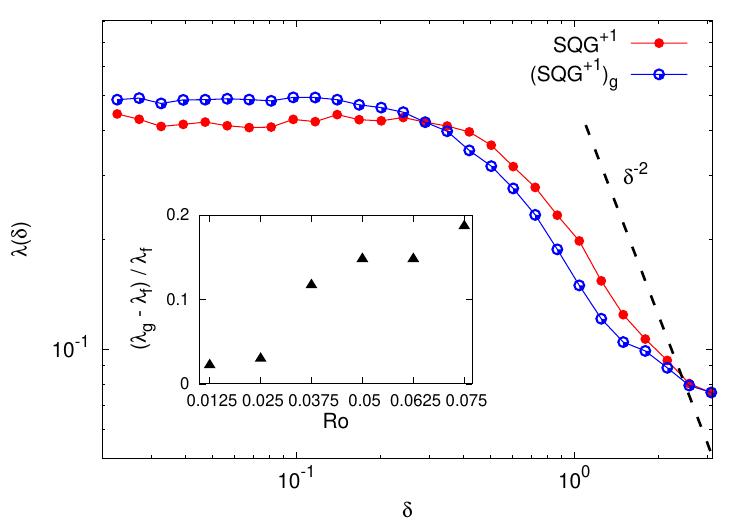}
\caption{
FSLE of the first kind $\lambda(\delta)$ for $Ro=0.0625$; 
colors and symbols have the same meaning as in Fig.~\ref{fig:vorticity_skewness}. 
The $\delta^{-2}$ scaling law is the dimensional expectation in the diffusive regime. 
Inset: relative difference between the plateau values (at small separations) of $\lambda(\delta)$ for particles advected by either the full flow or by its geostrophic component, as a function of $Ro$.
}
\label{fig:FSLE-I}
\end{figure}

Our measurement of $\lambda(\delta)$ is reported in Fig.~\ref{fig:FSLE-I} for $Ro=0.0625$, for both advection by the full and 
geostrophic-only flows. The results confirm those from $ku(t)$: dispersion is essentially nonlocal [$\lambda(\delta) \simeq \mathrm{const}$] over a broad range of separations, and the corresponding plateau value (an estimate of $\lambda_L$) is larger for advection by the geostrophic part of the flow only. 
This result also qualitatively agrees with the expectation that particle convergence, due to ageostrophic motions, reduces the dispersion rate. At the largest separations, the FSLE approaches the diffusive $\delta^{-2}$ scaling. The transition to this regime occurs at a smaller separation in the geostrophic-only flow, which appears consistent with the slightly smaller size of the largest eddies in this flow (see Sec.~\ref{sec:eulerian}).
Qualitatively similar results are found for the other Rossby numbers considered. 
 From a quantitative point of view, the differences due to filtering out ageostrophic velocities are quite small. However, the overestimation of the small-scale dispersion rate [the plateau value $\lambda(\delta) \simeq \mathrm{const}$] is not always negligible. 
Indeed, in the inset of Fig.~\ref{fig:FSLE-I}, we see that the relative difference $(\lambda_g-\lambda_f)/\lambda_f$ between those values computed in the full ($\lambda_f$) and geostrophic ($\lambda_g$) flow advection cases, grows monotonically and can reach about $20\%$ at the highest values of $Ro$. 
This finding appears relevant for Lagrangian dispersion applications relying on advection of synthetic drifters using real data from satellite altimetry, as the latter measures the geostrophic flow. Moreover, in real oceanic conditions the Rossby number 
could be even larger than in the present simulations, and thus this type of effects may be expected to be even more important. 

Most often, Eq.~(\ref{eq:FSLE}) is used to characterize the growth of the separation between two particles starting from different initial positions and evolving in the same flow. In such a case, $\lambda(\delta)$ is known as the FSLE of the first kind (FSLE-I). 
Another possibility is to apply the same computation to pairs of particles that start from the same position but evolve in two different flows, such as a reference flow and a perturbed one. This gives the FSLE of the second kind (FSLE-II), $\tilde{\lambda}(\delta)$, which is sometimes used to quantify the effect of unresolved flow components~\cite{boffettaetal2000,lacorata2019}.
Initially, particles start from the same position, hence the early growth of their distance is solely controlled by the differences in the velocity fields they are advected with. 
When their distance has sufficiently grown, the spatial increment of the velocity field will also contribute to their separation, and eventually dominate. 
This means that at large enough separations $\tilde{\lambda}(\delta)$ should approach $\lambda(\delta)$, while at small enough ones, the two kinds of FSLE should differ. This yields an estimate of a critical separation scale above which the flow perturbation has no significant effect on particle dynamics. 

\begin{figure}[htbp]
\centering
\includegraphics[width=0.6\textwidth]{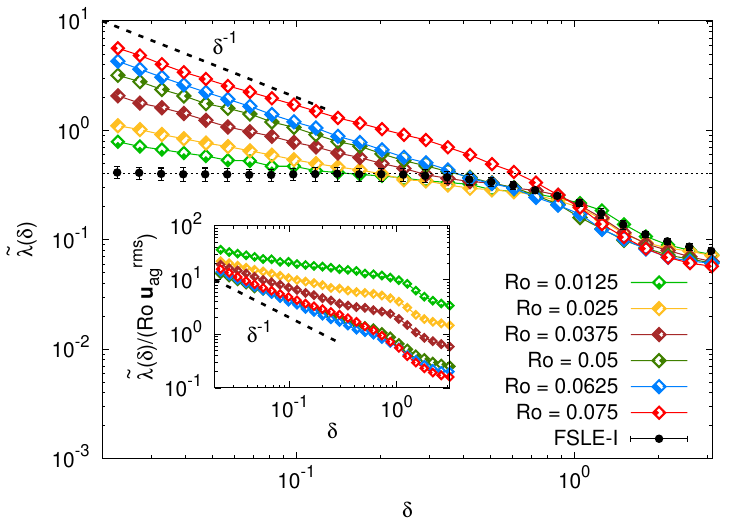}
\caption{FSLE of the second kind $\tilde{\lambda}(\delta)$ for different Rossby numbers. 
The filled black points correspond to the FSLE-I $\lambda(\delta)$, averaged over the values obtained at the different Rossby numbers. The uncertainty is here quantified by the standard deviation computed using the latter values. 
Inset: FSLE-II rescaled by the Rossby number and the rms value of the ageostrophic velocity.}
\label{fig:FSLE-II}
\end{figure}

Based on the above reasoning, we computed the FSLE of the second kind to provide a statistical characterization of the scale-dependent dispersion between trajectories computed from the full flow and from its geostrophic-only component 
The results are shown in Fig.~\ref{fig:FSLE-II}, for all the Rossby numbers explored. The filled black points are the average of the $\lambda(\delta)$ values obtained for different $Ro$ (which vary only weakly when such a control parameter is changed). As it can be seen, at large enough separations $\tilde{\lambda}(\delta)$ recovers the behavior of $\lambda(\delta)$, while at small ones it deviates from it to approach a $\delta^{-1}$ scaling. In this range of $\delta$ values, the role of the ageostrophic flow components, when present, is non negligible.

The behavior of the FSLE-II illustrated above can be explained as follows. 
First, recall that particle dynamics in the full and geostrophic-only flow are governed by $\dot{\bm{x}}=\bm{u}(\bm{x}(t),t)=\bm{u}_g(\bm{x}(t),t)+Ro\,\bm{u}_{ag}(\bm{x}(t),t)$ and  $\dot{\bm{x}}_g=\bm{u}_g(\bm{x}_g(t),t)$, respectively. Here, $\bm{x}(t)$ is the position of one of the two particles in a pair, advected by the total velocity, and $\bm{x}_g(t)$ is that of the other particle in the pair, advected only by the geostrophic velocity.
The particle separation vector $\Delta \bm{x}=\bm{x}-\bm{x}_g$ then evolves according to
\begin{equation}
\frac{d\Delta\bm{x}(t)}{dt}=
\bm{u}(\bm{x},t) - \bm{u}_g(\bm{x}_g,t).
\label{eq:sep_dyn}
\end{equation}
Adapting a more general derivation~\cite{boffettaetal2000} to our case, we 
perform a Taylor expansion of $\bm{u}(\bm{x},t)$ around $\bm{x}_g$:
\begin{equation}
\begin{split}
\bm{u}(\bm{x},t) & \simeq \bm{u}_g(\bm{x}_g,t) 
+ \left( \frac{\partial \bm{u}_g}{\partial \bm{x}} \right)_{\bm{x}_g} \Delta \bm{x} \\
& + Ro \left[ \bm{u}_{ag}(\bm{x}_g,t) + 
\left( \frac{\partial \bm{u}_{ag}}{\partial \bm{x}} \right)_{\bm{x}_g}\Delta \bm{x} \right],\\
&
\end{split}
\label{eq:full_vel_taylor}
\end{equation}
which implies
\begin{equation}
\begin{split}
\frac{d\Delta\bm{x}(t)}{dt} & \simeq 
\left( \frac{\partial \bm{u}_g}{\partial \bm{x}} \right)_{\bm{x}_g} \Delta \bm{x} \\ 
& + Ro \left[ \bm{u}_{ag}(\bm{x}_g,t) + 
\left( \frac{\partial \bm{u}_{ag}}{\partial \bm{x}} \right)_{\bm{x}_g} \Delta \bm{x} \right].\\
& 
\end{split}
\label{eq:sep_dyn_taylor}
\end{equation}
Since particles start from the same position [i.e. $\Delta \bm{x}(t_0)=0$], at short times Eq.~(\ref{eq:sep_dyn_taylor}) gives
\begin{equation}
\frac{d\Delta\bm{x}(t)}{dt} \simeq Ro \, \bm{u}_{ag}(\bm{x}_g,t).  
\label{eq:sep_dyn_early}
\end{equation}
From Eq.~(\ref{eq:sep_dyn_early}), 
using dimensional considerations, one has $\delta/t \sim Ro \, u_{ag}^{\mathrm{rms}}$ (with $u_{ag}^{\mathrm{rms}}=\sqrt{\langle |\bm{u}_{ag}|^2 \rangle}$). 
Therefore, the FSLE-II is expected to scale as
\begin{equation}
\tilde{\lambda}(\delta) \sim \frac{Ro \, u_{ag}^{\mathrm{rms}}}{\delta}.  
\label{eq:fsle-ii}
\end{equation}
As shown in the inset of Fig.~\ref{fig:FSLE-II}, the different curves 
are in fairly good agreement with the prediction in Eq.~(\ref{eq:fsle-ii}), except at the smallest nonzero Rossby number, and collapse onto each other 
for $Ro\ge0.05$.
At larger times, the separation distance $\Delta \bm{x}$ is no longer negligible and, eventually, the terms in $\Delta \bm{x}$ on the right-hand side of Eq.~(\ref{eq:sep_dyn_taylor}) dominate. For such large relative displacements, the particle separation distance evolves as if both particles were in the same flow, $d\Delta \bm{x}/dt \simeq (\partial_{\bm{x}} \bm{u} )_{\bm{x}_g} \Delta \bm{x}$. As a consequence, for large values of $\delta$ one finds $\tilde{\lambda}(\delta) \simeq \lambda(\delta)$,
as observed in Fig.~\ref{fig:FSLE-II}. 

The critical relative displacement $\delta^*$ below which the FSLE-II differs from the FSLE-I is found to increase with $Ro$. At the largest value of the latter ($Ro = 0.075$), we have $\tilde{\lambda}(\delta) \neq \lambda(\delta)$ over all separations, except in the diffusive range. 
If we exclude the data for $Ro=0.0125$, 
we observe that when $Ro$ increases from $0.025$ to $0.075$, i.e. by a factor $3$, $\delta^*$ increases from approximately $0.15$ to $0.8$, i.e. by a factor of roughly $5$. In spite of the idealized character of the present model dynamics, such values suggest caution when performing synthetic-particle advection, in the submesoscale range, with velocity fields derived from satellite altimetry. Indeed, the bias on the simulated trajectories, in terms of distance from the true ones, may be considerable given the possibly larger Rossby numbers of real ocean submesoscales with respect to those assumed here.

\subsection{Small-scale particle dynamics and clustering}
\label{sec:clustering}

In the previous section, we analyzed the separation process of Lagrangian tracers. However, through the metrics previously used it is not possible to address the quantitative characterization of aggregation phenomena.
For instance, the FSLE of the first kind (Fig.~\ref{fig:FSLE-I}) provides an estimate of the (scale-dependent) pair separation rate, but it does not allow to explore particle convergence events. 
Now, we investigate the small-scale particle dynamics for varying Rossby number, focusing on this aspect. This will also allow us to characterize particle clustering. 

An interesting tool to address this problem is offered by the spectrum of (asymptotic) Lyapunov exponents $\lambda_{1,2}$, with $\lambda_1 \geq \lambda_2$. 
These exponents can be computed by linearizing Eq.~(\ref{eq:motiontracers_f}) in tangent space and are thus related to the velocity gradient tensor (see  Appendix~\ref{app:lyap} and~\cite{lapeyre2002,CCV2010}).
While $\lambda_1$ measures the exponential divergence rate (and is positive for a chaotic system), $\lambda_2$ accounts for the dynamics along the local contracting direction. 
As the sum of Lyapunov exponents gives the divergence of the flow, $\lambda_1+\lambda_2=\bm{\nabla} \cdot \bm{u}$, clearly for an incompressible flow it is enough to compute $\lambda_1$. However, this is no longer the case in the presence of nonzero compressibility, as in our SQG$^{+1}$ simulations.
In such a case, it is instructive to separate the Lyapunov exponents into their contributions from nondivergent (or straining) and divergent processes. To this end, we introduce $s=\lambda_1-\lambda_2$ and $d=\lambda_1+\lambda_2$, so that $\lambda_1=(s+d)/2$ and $\lambda_2=(-s+d)/2$. Since we know that the SQG$^{+1}$ flow is turbulent, with particle pair separations eventually increasing in time, $\lambda_1$ should be positive. 
Due to the occurrence of clustering at small scales, we also expect $d \leq 0$, implying that $|\lambda_2| \geq \lambda_1$ and $s>0$. 
Then, from the expressions of $\lambda_1$ and $\lambda_2$ it is possible to see that both Lyapunov exponents should be reduced by the nonzero divergence, with respect to those of the incompressible part of the flow. 

\begin{figure}[htbp]
\centering
\includegraphics[width=0.49\textwidth]{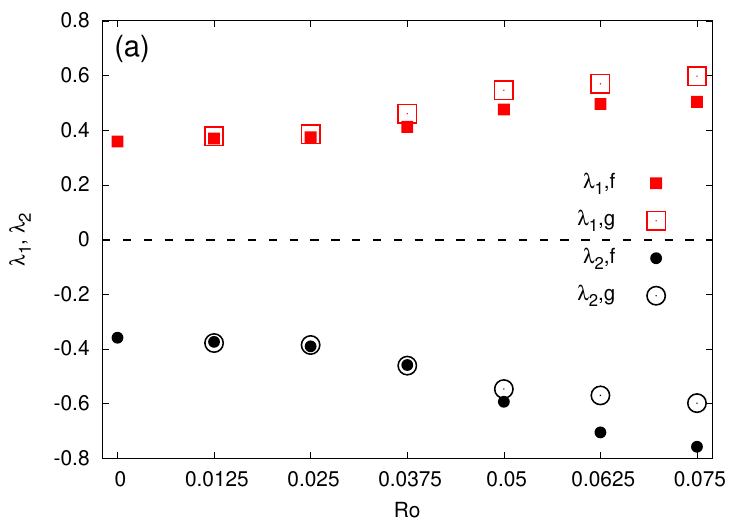}
\includegraphics[width=0.49\textwidth]{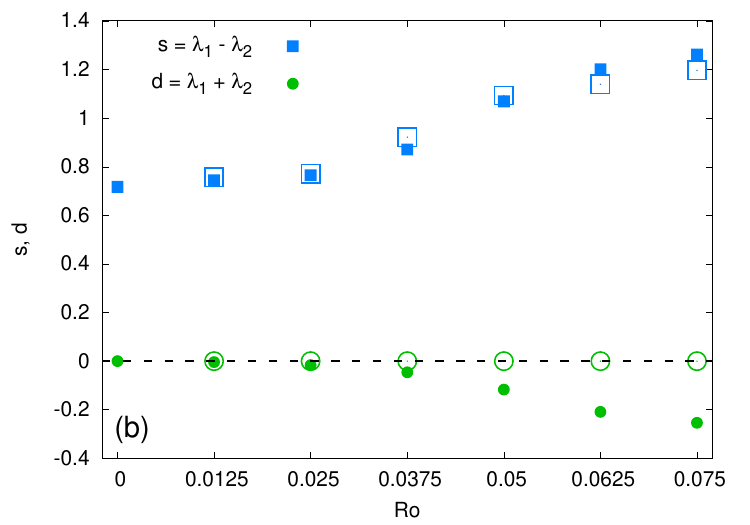}
\caption{(a) Lyapunov exponents 
of the Lagrangian dynamics (red squares for $\lambda_1$ and black circles for $\lambda_2$) for varying Rossby number.   
(b) Mean Lagrangian strain, $s=\lambda_1-\lambda_2$, and divergence, $d=\lambda_1+\lambda_2$ versus $Ro$.
In both (a) and (b) the filled and empty points are calculated from SQG$^{+1}$ flows and their geostrophic component, respectively. Uncertainties, estimated as the standard deviation of $\lambda_{1,2}$ over their time series (at large times), are mostly of the order of point size.}
\label{fig:lyapunov_spectrum}
\end{figure}

The Lyapunov exponents computed using the full and 
geostrophic-only flows are shown in Fig.~\ref{fig:lyapunov_spectrum}a as a function of the Rossby number {(see Appendix~\ref{app:lyap} and~\cite{benettinetal1980,lapeyre2002,CCV2010} for more details on the computation method)}. 
Here, we also present the values obtained in the simulation of SQG turbulence (i.e. for $Ro=0$).
The values of $d=\lambda_1+\lambda_2$ and $s=\lambda_1-\lambda_2$ versus $Ro$ are shown in Fig.~\ref{fig:lyapunov_spectrum}b 
[in both panels (a) and (b) an average over all the different Lagrangian initial conditions is also taken].
As expected, for $Ro=0$, the two Lyapunov exponents sum to zero, $\lambda_2(0)=-\lambda_1(0)$ [$d(0)=0$].
For nonzero and increasing $Ro$, both $\lambda_{1,f}$ and $\lambda_{2,f}$ grow in absolute value, but $\lambda_{2,f}$ by a larger amount, so that $|\lambda_{2,f}| > \lambda_{1,f}$ (i.e. $\lambda_{1,f}+\lambda_{2,f}<0$) at all $Ro$ (here the subscript $f$ indicates that the full flow is considered). 
The mean Lagrangian divergence $d$ (the average being over particles) is consistently negative, growing in absolute value with $Ro$ (Fig.~\ref{fig:lyapunov_spectrum}b). 
In the (SQG$^{+1}$)$_g$ case, the flow is nondivergent by construction, because only the geostrophic velocity component is retained. As it can be seen in Fig.~\ref{fig:lyapunov_spectrum}b this constraint is very well satisfied in our simulations. 
The mean Lagrangian strain $s$ does not differ much between the (SQG$^{+1}$) and (SQG$^{+1}$)$_g$ cases, i.e. $s_f\simeq s_g$ (the subscript $g$ indicating that the geostrophic-only flow is considered) for all Rossby numbers. This implies that filtering out $\bm{u}_{ag}$ only affects the divergent part of velocity gradients and much less the straining one. 
Since $s_f(Ro)\approx s_g(Ro)$ and $d_f(Ro)\le0$, we have
$\lambda_{1,g}(Ro)=s_g(Ro)/2\approx s_f(Ro)/2\geq[s_f(Ro)+d_f(Ro)]/2=\lambda_{1,f}(Ro)$ and 
$\lambda_{2,g}(Ro)=-s_g(Ro)/2\approx -s_f(Ro)/2\geq[-s_f(Ro)+d_f(Ro)]/2=\lambda_{2,f}(Ro)$. 
This explains why $\lambda_{i,g}(Ro) \geq \lambda_{i,f}(Ro)$ (with $i=1,2$), as observed in Fig.~\ref{fig:lyapunov_spectrum}a.
These arguments then provide support to the increased FSLE-I plateau value in the (SQG$^{+1}$)$_g$ case (Fig.~\ref{fig:FSLE-I}).
Note that we verified that the values of $[\lambda_{1,g}(Ro)-\lambda_{1,f}(Ro)]/\lambda_{1,f}(Ro)$ nicely match those of the FSLE-I relative difference {(at not too large separations)} shown in the inset of Fig.~\ref{fig:FSLE-I}.
In addition, these results indicate, once more, that filtering the SQG$^{+1}$ flow to exclude ageostrophic motions does not lead to the same flow properties as those of a purely SQG system (i.e. with $Ro=0$).

Lyapunov exponents also provide further information on the clustering of Lagrangian tracers. In particular, they can be used to compute the fractal dimension of the sets over which particles accumulate (when the full flow is considered). This is known as the Lyapunov dimension~\cite{CCV2010}, and in the present 2D case it is given by
\begin{equation}
{D_L=1+\frac{\lambda_1}{|\lambda_2|}.}
\label{eq:DL}
\end{equation}
Note that for an incompressible flow (like the geostrophic one) one would have $\lambda_1 = |\lambda_2|$, and hence $D_L=2$, meaning uniformly distributed particles. 
As in SQG$^{+1}$ the geostrophic equilibrium is broken and the flow becomes compressible, $|\lambda_2| > \lambda_1$ and $D_L<2$, implying particle clustering. From Eq.~(\ref{eq:DL}), when $|\lambda_2| \gg \lambda_1$ one has $D_L \simeq 1$, i.e. a one-dimensional (1D) fractal set. 
Clustering is clearly due to the compressibility of the horizontal flow being nonzero, and in the following we will thus discuss the relation between $D_L$ and this quantity. 
However, the flow compressibility alone typically does not allow to fully characterize the distribution of particles~\cite{cressmanetal2004,boffettaetal2004}. Different other factors can be also important and, among these, the flow time correlations play a relevant role~\cite{boffettaetal2004,dhanagareetal2014}, as we shall see below for our system.

\begin{figure}[htbp]
\centering
\includegraphics[width=0.6\textwidth]{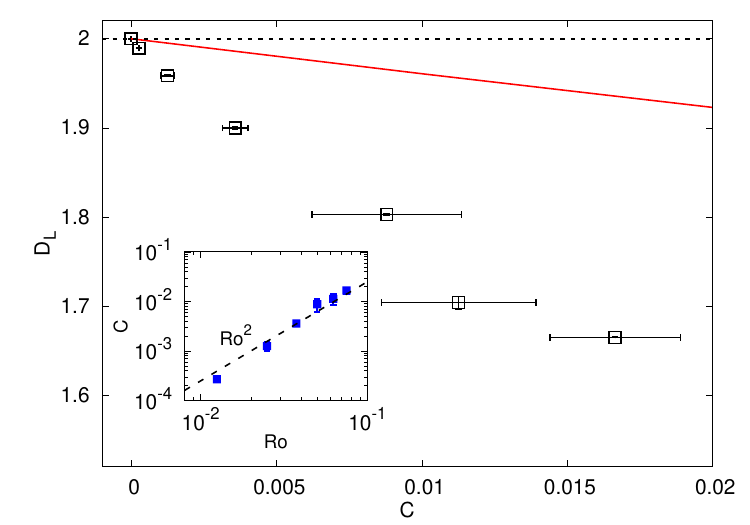}
\caption{ 
Lyapunov dimension $D_L$, {for particles advected by the full flow,} as a function of compressibility $\mathcal{C}$; the solid red line is the expectation $D_L=2/(1+2\mathcal{C})$
in the compressible Kraichnan model. 
Inset: compressibility versus $Ro$ and the prediction $\mathcal{C} \sim Ro^2$ (dashed line).
Uncertainties on $D_L$ and $\mathcal{C}$ are estimated from the standard deviation using the values taken over their time series (at large times).
}
\label{fig:fract_dim}
\end{figure}

The compressibility of the (full) Eulerian flow is quantified by the ratio~\cite{cressmanetal2004,boffettaetal2004,dhanagareetal2014} 
\begin{equation}
\mathcal{C}=\frac{\langle (\partial_x u + \partial_y v)^2\rangle}{\langle (\partial_x u)^2 + (\partial_x v)^2 + (\partial_y u)^2  + (\partial_y v)^2\rangle}, 
\label{eq:compressibility_def}
\end{equation}
which takes values between $0$ and $1$, for incompressible and potential flow, respectively. 
Providing a theoretical prediction for  $\mathcal{C}$ from its definition is generally not an easy task as it requires estimating the correlations of velocity gradients. 
Indeed, the denominator in Eq.~(\ref{eq:compressibility_def}) can be rewritten as $\langle \Delta^2 \rangle + \langle \zeta^2 \rangle -2 \langle (\partial_x u) \partial_y v - (\partial_x v) \partial_y u \rangle$, where 
the correlations between different velocity-gradient components are more evident,
$\Delta$ is divergence and $\zeta$ is vorticity. 
The structure of the velocity-gradient tensor and its low-order moments, were recently analyzed for both incompressible~\cite{pumir2017} and compressible~\cite{yangetal2022} three-dimensional (3D) turbulence. 
Using the same derivation as in~\cite{yangetal2022} and under the assumptions of homogeneity and isotropy, we obtain in the 2D case $\langle (\partial_x u) \partial_y v \rangle = \langle (\partial_x v) \partial_y u \rangle$. 
This relation is found to be well verified in our simulations for all Rossby numbers (see Appendix~\ref{app:compr_ratio}). 
Compressibility is then given by
\begin{equation}
\mathcal{C}=\frac{\langle \Delta^2\rangle}{\langle \Delta^2 \rangle + \langle \zeta^2 \rangle}. 
\label{eq:compressibility}
\end{equation}
Considering now that $\bm{u}=\bm{u}_g + Ro \, \bm{u}_{ag}$, one has $\Delta = \bm{\nabla} \cdot \bm{u}=Ro \, \bm{\nabla} \cdot \bm{u}_{ag}$ and $\zeta=\zeta_g + Ro \, \zeta_{ag}$. 
Inserting these expressions in Eq.~(\ref{eq:compressibility}), at lowest order we then obtain the following estimate of $\mathcal{C}$ as a function of $Ro$
\begin{equation}
\mathcal{C}=\frac{Ro^2}{Ro^2 + 1} \sim Ro^2.  
\label{eq:compressibility_scaling}
\end{equation}
As seen in the inset of Fig.~\ref{fig:fract_dim}, our numerical data are in quite good agreement with Eq.~(\ref{eq:compressibility_scaling}), supporting this prediction.

While here compressibility is always small, due to Eq.~(\ref{eq:compressibility_scaling}), 
clustering is well evident in the SQG$^{+1}$ system, as highlighted by the decrease of $D_L$ with $\mathcal{C}$ (Fig.~\ref{fig:fract_dim}). 
For the SQG case ($Ro=0$ and $\mathcal{C}=0$), the Lyapunov dimension is very close to $2$, in agreement with the nondivergent nature of this flow.
As $Ro$ (and then also $\mathcal{C}$) grows, it decreases monotonically and its value allows to quantify the intensity of clustering. 
Such decrease is due to $|\lambda_{2,f}|$ growing faster with $Ro$ than $\lambda_{1,f}$ (Fig.~\ref{fig:lyapunov_spectrum}a), meaning to the intensification, and dominance, of the locally contracting flow direction. 
These findings indicate that the structures over which particles accumulate are not space-filling, and tend to be more and more unidimensional for larger $Ro$. 
This in turn suggests that clustering should occur  along filaments, which is in line with the observations from Fig.~\ref{fig:vorticity_diff}c. 
By filtering the flow to take only its geostrophic component, instead, with good accuracy we retrieve $D_L = 2$ (not shown), corresponding to particles filling the entire domain (see also Fig.~\ref{fig:vorticity_diff}d). 

On the basis of the persistent structures present in our flows (see Sections~\ref{sec:eulerian} and~\ref{sec:lagr_disp}),
we argue that the relevant decrease of $D_L$, in spite of the small compressibility, is due to the time correlations in the velocity field. To test this hypothesis, we compare our results with what one would obtain in a temporally uncorrelated flow. For this purpose, we consider the 2D compressible Kraichnan flow, which is white-in-time, and for which the following prediction~\cite{boffettaetal2004,dhanagareetal2014} for $D_L$ is available:
\begin{equation}
D_L=\frac{2}{1+2\mathcal{C}} .
\label{eq:kraichnan_DL}
\end{equation}
Figure~\ref{fig:fract_dim}, where the Kraichnan-model prediction is the solid red line, shows that in the absence of flow temporal correlations the fractal dimension is considerably larger than in the SQG$^{+1}$ system. This indicates that in the present case clustering is essentially due to the interplay between the (small) Eulerian compressibility and the existence of long-lived flow structures that trap particles, enhancing their aggregation. 
We note that this finding may also be understood by considering the evolution equation for the 
gradients of the particle density field $\rho_p(\bm{x},t)$. The latter is defined as the number of Lagrangian tracers per area and is governed by the equation $\partial_t \rho_p + \bm{u} \cdot \bm{\nabla} \rho_p=-\rho_p \bm{\nabla} \cdot
\bm{u}$. For the gradients of $\rho_p$ one then has: 
\begin{equation}
\frac{D}{Dt} \bm{\nabla} \rho_p = 
- \left(\bm{\nabla} \bm{u}\right)^T \bm{\nabla} \rho_p
-\Delta \bm{\nabla} \rho_p
- \rho_p \bm{\nabla} \Delta,
\label{eq:scalar_grad}
\end{equation}
in which 
$D/Dt=\partial_t + \bm{u} \cdot \bm{\nabla}$, $\left(\bm{\nabla} \bm{u}\right)^T$ denotes the transpose of the velocity gradient tensor and $\Delta$ is divergence. 
From Eq.~(\ref{eq:scalar_grad}), one can see that 
$\Delta$, which is due to ageostrophic corrections, 
if nonuniform ($\bm{\nabla} \Delta \neq 0$), will create gradients of the scalar field $\rho_p$ that will be  amplified in convergence regions ($\Delta<0$), and further strengthened by persistent strain [related to the velocity-gradient tensor $\left(\bm{\nabla} \bm{u}\right)$]. 
Note, too, that it is not difficult to realize that rotation does not affect the magnitude of $\bm{\nabla} \rho_p$. 
As strain is related to the structure of the flow, in the SQG$^{+1}$ case its persistence reflects the time correlations of the velocity field. In the temporally uncorrelated Kraichnan flow, instead, this effect is not present, which leads to weaker clustering.

We conclude by noting that the transition to strong clustering, with particles accumulating over 1D structures, is marked by the Lyapunov dimension reaching $D_L=1$. This occurs for a critical compressibility $\mathcal{C}^*=1/2$ in Kraichnan model. Based on the results in Fig.~\ref{fig:fract_dim}, with the numerical data being always below the theoretical prediction of Eq.~(\ref{eq:kraichnan_DL}), one may speculate that in the SQG$^{+1}$ system, the transition occurs for $\mathcal{C}^*<1/2$. 
From $\mathcal{C}^*$, the corresponding critical Rossby number may then be estimated as $Ro^* \approx {\mathcal{C}^*}^{1/2}$.  
However, clustering properties in time-correlated compressible flows strongly depend on the spatio-temporal details of the velocity field~\cite{dhanagareetal2014}. Indeed, it has been shown that for Lagrangian tracers at the free surface of a 3D incompressible Navier-Stokes turbulent flow~\cite{boffettaetal2004}, 
while the qualitative behavior of $D_L$ as a function of $\mathcal{C}$ is similar to that observed here for small $\mathcal{C}$, the transition occurs at $\mathcal{C}^*>1/2$. 
The determination of the critical compressibility (and Rossby number) for SQG$^{+1}$ turbulence thus remains an open question, which would require considerably extending the range of $Ro$ values explored and extensive numerical simulations.

\section{Conclusions}\label{sec:conclusion}

We investigated surface-ocean turbulence in the fine-scale range by means of numerical simulations of the SQG$^{+1}$ model~\cite{HSM2002,maaloulyetal2023}. This model is derived from primitive equations and extends the SQG one by including ageostrophic motions corresponding to first-order corrections in the Rossby number. By construction the latter are related to secondary flows due to finite-Rossby effects at fronts. Note that other ageostrophic processes (as, e.g., internal waves), further deviating from geostrophy, are not represented~\cite{sinha19}.  
As previously shown~\cite{maaloulyetal2023}, 
this approach allows to reproduce both a prevalence of cyclones over anticyclones (see also~\cite{du2024}) and the accumulation of Lagrangian tracers in cyclonic frontal regions, which are found in observations~\cite{Rudnick2001,Shcherbina_etal_2013,Dasaro_etal_2018,Jacobs_etal_2016,Berta_etal_2020} but not captured by standard QG models. Our main goal was to assess the effect of ageostrophic motions on Lagrangian pair dispersion, which is relevant for the interpretation and exploitation of new, high-resolution satellite data~\cite{morrow2023ocean,fu2024surface}, as well as to improve the understanding of material spreading at the surface of the ocean. For this purpose we compared Lagrangian statistics for tracer particles advected by either the full SQG$^{+1}$ flow or by its 
geostrophic component, for different Rossby numbers.

Our results confirm the general expectation, also supported by previous numerical indications~\cite{maaloulyetal2023}, that relative dispersion weakly depends on the ageostrophic corrections to the flow. From a quantitative point of view, however, the FSLE-I, a fixed lengthscale indicator of the separation process, reveals that excluding the ageostrophic velocity in the advection leads to an overestimation of the typical pair-dispersion rate, and that the importance of this effect grows with $Ro$. This can be understood by analyzing the spectrum of the (asymptotic) Lyapunov exponents of the particle dynamics. Considering the weak dependence of the FSLE-I on spatial scales in the present simulations, the latter appear appropriate to characterize the small-scale behavior of particles over a significant range of scales. A decomposition of Lyapunov exponents into the divergent and nondivergent parts of the velocity-gradient tensor experienced by particles shows that the absence of flow convergences in the geostrophic-only case is at the origin of the increase of both exponents, and hence of the FSLE-I at the smallest separations.

In addition, we examined the scale-by-scale dispersion rate for pairs such that both particles start from the same position but one evolves in the full flow and the other in its geostrophic component only.
We found that such an inter-model dispersion rate (FSLE-II) differs from the FSLE-I over a range of small separations, which extends towards larger and larger ones with $Ro$. 
The behavior of the FSLE-II is explained by a simple theoretical argument relying on the different mechanisms (the differences in the evolution equations and in the particle positions) controlling the separation process. 
These results highlight that, at sufficiently small separations, 
particle trajectories are sensitive to ageostrophic motions and can be biased if advected by the geostrophic velocity only, which appears relevant to applications using satellite-derived velocity fields to advect synthetic particles in order to deduce flow transport properties.

Beyond the above quantitative differences, the ageostrophic velocities are responsible of a major qualitative change in the Lagrangian dynamics, namely the occurrence of clustering of tracer particles. While this is clearly not captured by geostrophic flows, which are incompressible by definition, it has important consequences for the identification of hotspots of pollutant accumulation in the sea and for marine-ecology modeling. We then measured its intensity for increasing Rossby numbers and characterized the mechanisms controlling it in the SQG$^{+1}$ system. We showed that the horizontal-flow compressibility is always small and grows only quadratically with $Ro$. Nevertheless, clustering can be relatively intense, with the Lyapunov dimension clearly decreasing to values smaller than 2 
with increasing $Ro$ (and compressibility). Finally, the comparison of our numerical results with the prediction for the time uncorrelated Kraichnan flow~\cite{cressmanetal2004,boffettaetal2004} revealed that clustering is, in the present case, essentially due to the interplay between the small compressibility and the important temporal correlations of the flow.

To conclude, this study indicates that the overall effect of ageostrophic motions related to fronts on Lagrangian pair dispersion at the ocean surface should be weak. Nevertheless, it also suggests some caution in particle advection experiments with geostrophically derived flows, as single-particle trajectories should separate from the true ones, and important phenomena, such as clustering, would be missed. 
An interesting perspective of this work would be to extend the analysis to realistic circulation models, in order to address the impact on Lagrangian dynamics of other ageostrophic processes (internal gravity waves and tides) that are associated with the ocean fast variability.

\section*{Acknowledgments}

{This work was supported by CNES, focused on SWOT space mission, in the framework of DIEGO and DIEGOB projects.}

\appendix
\section{Lyapunov exponents' spectrum}\label{app:lyap}

Lyapunov's theory of dynamical systems~\cite{CCV2010} can be applied to the evolution equation of Lagrangian tracer particles
\begin{equation}
    \frac{d\bm{x}}{dt}=\bm{u}(\bm{x}(t),t).
    \label{eq:dynsyst}
\end{equation}
The linearized version of Eq.~(\ref{eq:dynsyst}), in tangent space, is just 
\begin{equation}
    \frac{d\bm{w}}{dt}=[\nabla\bm{u}](\bm{x}(t),t)~\bm{w}.
    \label{eq:tangent_syst}
\end{equation}
The above equation is integrated along the Lagrangian path $\bm{x}(t)$; here $[\nabla\bm{u}](\bm{x}(t),t)$ is the velocity gradient tensor at position $\bm{x}$ at time $t$. 
Equation~(\ref{eq:tangent_syst}) can be viewed as the equation for the separation $\delta\bm{x}$ between two (infinitesimally) close Lagrangian trajectories~\cite{lapeyre2002}. 

The Lyapunov spectrum is related to the asymptotic exponential growth rate of $\bm{w}$ and is computed as follows~\cite{benettinetal1980,CCV2010}.
Given an arbitrary unitary initial vector
$\bm{w}_1(t_0)$, the first exponent is computed as
\begin{equation}
\lambda_1=\lim_{t \rightarrow \infty} \frac{1}{t-t_0} \ln{\left( \frac{|\bm{w}_1(t)|}{|\bm{w}_1(t_0)|} \right)},
\label{eq:lambda1}
\end{equation}
The second exponent is computed through the use of a second vector $\bm{w}_2(t)$, initially unitary and orthogonal to the first one, evolving according to the same equation. The area $A(t)$ of the parallelogram defined by  $\bm{w}_1(t)$ and $\bm{w}_2(t)$ at each time $t$ allows to introduce $\Lambda$ such that
\begin{equation}
\Lambda=\lim_{t \rightarrow \infty} \frac{1}{t-t_0} \ln{\left[ \frac{A(t)}{A(t_0)} \right]}.
\label{eq:Lambda}
\end{equation}
Once $\Lambda$ is known, $\lambda_2$ can be computed as
\begin{equation}
\lambda_2=\Lambda-\lambda_1,
\label{eq:lambda2}
\end{equation}
More details about the implementation of this method can be found in~\cite{CCV2010}. 
Note that using an ensemble of particles, we obtain values of $\lambda_i$ ($i=1$, $2$) for each trajectory, which should be the same assuming ergodicity. In practice, $\lambda_i$ values are further averaged over all trajectories.

\section{Compressibility ratio}\label{app:compr_ratio}

{The compressibility ratio of Eq.~(\ref{eq:compressibility_def}), $\mathcal{C}=\langle (\bm{\nabla} \cdot \bm{u})^2\rangle/\langle (\bm{\nabla}  \bm{u})^2\rangle$, accounts for the relative strength of divergence and strain. Considering that 
$$
(\bm{\nabla}  \bm{u})^2=(\partial_x u)^2 + (\partial_x v)^2 + (\partial_y u)^2 + (\partial_y v)^2,
$$
$$
\Delta^2 \equiv (\bm{\nabla} \cdot \bm{u})^2=(\partial_x u)^2 + (\partial_y v)^2 + 2 \, \partial_x u \, \partial_y v,
$$ 
$$
\zeta^2 = (\partial_x v)^2 + (\partial_y u)^2 - 2 \, \partial_x v \, \partial_y u,
$$
one has $\Delta^2+\zeta^2=(\bm{\nabla}  \bm{u})^2 + 2 (\partial_x u \, \partial_y v-\partial_x v \, \partial_y u)$. Therefore, the compressibility ratio can be also written as
\begin{equation}
\mathcal{C} = \frac{\langle \Delta^2 \rangle}{\langle \Delta^2 \rangle + \langle \zeta^2 \rangle - 2 \left(\langle \partial_x u \, \partial_y v \rangle - \langle \partial_x v \, \partial_y u \rangle\right)}.
\label{eq:compr_ratio_1}
\end{equation}
}

\begin{figure}[htbp]
\centering
\includegraphics[width=0.6\textwidth]{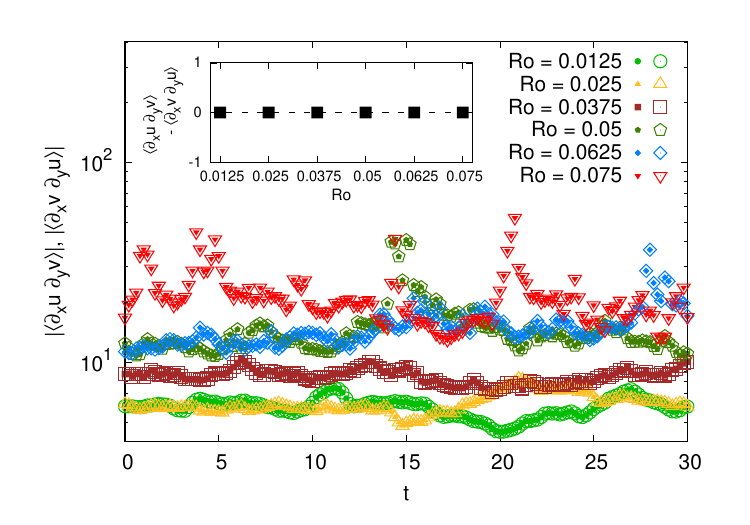}
\caption{{Velocity-gradient correlations $\langle \partial_x u \, \partial_y v \rangle$ (filled points) and $\langle \partial_x v \, \partial_y u \rangle$ (empty points) as a function of time for the full SQG$^{+1}$ turbulent flow and different Rossby numbers (different point types). Inset: $\langle \partial_x u \, \partial_y v \rangle - \langle \partial_x v \, \partial_y u \rangle$, temporally averaged in the statistically steady state of the flow, as a function of the Rossby number. 
}
}
\label{fig:velgrad_correl}
\end{figure}

{To further simplify Eq.~(\ref{eq:compr_ratio_1}), one needs to estimate the correlations of velocity gradients appearing in the last parenthesis in the denominator. 
This problem was addressed in~\cite{yangetal2022} in a broader context, to characterize the low-order moments of velocity gradients of 3D compressible flows. Here we recall some of the main points of the reasoning, adapting them to our 2D case. 
Specifically, we define $A^{(2)}_{ijkl}=\langle \partial_j u_i \, \partial_l u_k \rangle$, where clearly $i,j,k,l=1,2$ (indices $1$ and $2$ corresponding to the $x$ and $y$ directions, respectively) in 2D. 
As shown in~\cite{yangetal2022}, assuming statistical homogeneity ($\partial_i \langle ... \rangle=0$) one has 
\begin{equation}
A^{(2)}_{ijji}=\langle \partial_j u_i \, \partial_i u_j \rangle = \langle \partial_i u_i \, \partial_j u_j \rangle = A^{(2)}_{iijj},
\label{eq:velgrad_corr_homo}
\end{equation}
where repeated indices are summed over. 
For isotropic flows the velocity-gradient correlation tensor can be expressed as
\begin{equation}
A^{(2)}_{ijkl}=\alpha \, \delta_{ij} \delta_{kl} + \beta \, \delta_{ik} \delta_{jl} + \gamma \, \delta_{il} \delta_{jk},
\label{eq:velgrad_corr_iso}
\end{equation}
with $\alpha$, $\beta$, $\gamma$ some constants and $\delta_{ij}$ indicating the Kronecker tensor. Using Eq.~(\ref{eq:velgrad_corr_iso}), one gets that $A^{(2)}_{ijji}=2\alpha+2\beta+4\gamma$ and $A^{(2)}_{iijj}=4\alpha+2\beta+2\gamma$, implying $\alpha=\gamma$ thanks to the constraint in Eq.~(\ref{eq:velgrad_corr_homo}). This last relation has the following important consequence:
\begin{equation}
\langle \partial_1 u_1 \, \partial_2 u_2 \rangle = A^{(2)}_{1122} = A^{(2)}_{1221} =\langle \partial_2 u_1 \, \partial_1 u_2 \rangle,
\label{eq:velgrad_corr_zero}
\end{equation}
since $A^{(2)}_{1122}=\alpha$ and $A^{(2)}_{1221}=\gamma$, from Eq.~(\ref{eq:velgrad_corr_iso}). Coming back to our original notation, this means that
\begin{equation}
\langle \partial_x u \, \partial_y v \rangle - \langle \partial_x v \, \partial_y u \rangle = 0.
\label{eq:velgrad_corr_zero_xy}
\end{equation}
The above relation is very well verified in our numerical simulations for all Rossby numbers (Fig.~\ref{fig:velgrad_correl}), and 
allows us to write the compressibility ratio as $\mathcal{C}=\langle \Delta^2\rangle/\left(\langle \Delta^2 \rangle + \langle \zeta^2 \rangle\right)$, i.e. as in Eq.~(\ref{eq:compressibility}).} 

\bibliography{references}

\begin{thebibliography}{58}%
\makeatletter
\providecommand \@ifxundefined [1]{%
 \@ifx{#1\undefined}
}%
\providecommand \@ifnum [1]{%
 \ifnum #1\expandafter \@firstoftwo
 \else \expandafter \@secondoftwo
 \fi
}%
\providecommand \@ifx [1]{%
 \ifx #1\expandafter \@firstoftwo
 \else \expandafter \@secondoftwo
 \fi
}%
\providecommand \natexlab [1]{#1}%
\providecommand \enquote  [1]{``#1''}%
\providecommand \bibnamefont  [1]{#1}%
\providecommand \bibfnamefont [1]{#1}%
\providecommand \citenamefont [1]{#1}%
\providecommand \href@noop [0]{\@secondoftwo}%
\providecommand \href [0]{\begingroup \@sanitize@url \@href}%
\providecommand \@href[1]{\@@startlink{#1}\@@href}%
\providecommand \@@href[1]{\endgroup#1\@@endlink}%
\providecommand \@sanitize@url [0]{\catcode `\\12\catcode `\$12\catcode `\&12\catcode `\#12\catcode `\^12\catcode `\_12\catcode `\%12\relax}%
\providecommand \@@startlink[1]{}%
\providecommand \@@endlink[0]{}%
\providecommand \url  [0]{\begingroup\@sanitize@url \@url }%
\providecommand \@url [1]{\endgroup\@href {#1}{\urlprefix }}%
\providecommand \urlprefix  [0]{URL }%
\providecommand \Eprint [0]{\href }%
\providecommand \doibase [0]{https://doi.org/}%
\providecommand \selectlanguage [0]{\@gobble}%
\providecommand \bibinfo  [0]{\@secondoftwo}%
\providecommand \bibfield  [0]{\@secondoftwo}%
\providecommand \translation [1]{[#1]}%
\providecommand \BibitemOpen [0]{}%
\providecommand \bibitemStop [0]{}%
\providecommand \bibitemNoStop [0]{.\EOS\space}%
\providecommand \EOS [0]{\spacefactor3000\relax}%
\providecommand \BibitemShut  [1]{\csname bibitem#1\endcsname}%
\let\auto@bib@innerbib\@empty
\bibitem [{\citenamefont {Vallis}(2017)}]{Vallis2017}%
  \BibitemOpen
  \bibfield  {author} {\bibinfo {author} {\bibfnamefont {G.~K.}\ \bibnamefont {Vallis}},\ }\href@noop {} {\emph {\bibinfo {title} {Atmospheric and {Oceanic} {Fluid} {Dynamics}}}}\ (\bibinfo  {publisher} {Cambridge University Press, New York, USA},\ \bibinfo {year} {2017})\BibitemShut {NoStop}%
\bibitem [{\citenamefont {Zhang}\ \emph {et~al.}(2014)\citenamefont {Zhang}, \citenamefont {Wang},\ and\ \citenamefont {Qiu}}]{zhang2014oceanic}%
  \BibitemOpen
  \bibfield  {author} {\bibinfo {author} {\bibfnamefont {Z.}~\bibnamefont {Zhang}}, \bibinfo {author} {\bibfnamefont {W.}~\bibnamefont {Wang}},\ and\ \bibinfo {author} {\bibfnamefont {B.}~\bibnamefont {Qiu}},\ }\bibfield  {title} {\bibinfo {title} {Oceanic mass transport by mesoscale eddies},\ }\href@noop {} {\bibfield  {journal} {\bibinfo  {journal} {Science}\ }\textbf {\bibinfo {volume} {345}},\ \bibinfo {pages} {322} (\bibinfo {year} {2014})}\BibitemShut {NoStop}%
\bibitem [{\citenamefont {Klein}\ and\ \citenamefont {Lapeyre}(2009)}]{klein2009oceanic}%
  \BibitemOpen
  \bibfield  {author} {\bibinfo {author} {\bibfnamefont {P.}~\bibnamefont {Klein}}\ and\ \bibinfo {author} {\bibfnamefont {G.}~\bibnamefont {Lapeyre}},\ }\bibfield  {title} {\bibinfo {title} {The oceanic vertical pump induced by mesoscale and submesoscale turbulence},\ }\href@noop {} {\bibfield  {journal} {\bibinfo  {journal} {Annu. Rev. Mar. Sci.}\ }\textbf {\bibinfo {volume} {1}},\ \bibinfo {pages} {351} (\bibinfo {year} {2009})}\BibitemShut {NoStop}%
\bibitem [{\citenamefont {Zhang}\ \emph {et~al.}(2019)\citenamefont {Zhang}, \citenamefont {Qiu}, \citenamefont {Klein},\ and\ \citenamefont {Travis}}]{ZQKT2019}%
  \BibitemOpen
  \bibfield  {author} {\bibinfo {author} {\bibfnamefont {Z.}~\bibnamefont {Zhang}}, \bibinfo {author} {\bibfnamefont {B.}~\bibnamefont {Qiu}}, \bibinfo {author} {\bibfnamefont {P.}~\bibnamefont {Klein}},\ and\ \bibinfo {author} {\bibfnamefont {S.}~\bibnamefont {Travis}},\ }\bibfield  {title} {\bibinfo {title} {The influence of geostrophic strain on oceanic ageostrophic motion and surface chlorophyll},\ }\href@noop {} {\bibfield  {journal} {\bibinfo  {journal} {Nat. Commun.}\ }\textbf {\bibinfo {volume} {10}},\ \bibinfo {pages} {2838} (\bibinfo {year} {2019})}\BibitemShut {NoStop}%
\bibitem [{\citenamefont {McWilliams}(2016)}]{McWilliams2016}%
  \BibitemOpen
  \bibfield  {author} {\bibinfo {author} {\bibfnamefont {J.~C.}\ \bibnamefont {McWilliams}},\ }\bibfield  {title} {\bibinfo {title} {Submesoscale currents in the ocean},\ }\href@noop {} {\bibfield  {journal} {\bibinfo  {journal} {Proc. R. Soc. A}\ }\textbf {\bibinfo {volume} {472}},\ \bibinfo {pages} {20160117} (\bibinfo {year} {2016})}\BibitemShut {NoStop}%
\bibitem [{\citenamefont {Morrow}\ \emph {et~al.}(2019)\citenamefont {Morrow}, \citenamefont {Fu}, \citenamefont {Ardhuin}, \citenamefont {Benkiran}, \citenamefont {Chapron}, \citenamefont {Cosme}, \citenamefont {{d'Ovidio}}, \citenamefont {Farrar}, \citenamefont {Gille}, \citenamefont {Lapeyre}, \citenamefont {{Le Traon}}, \citenamefont {Pascual}, \citenamefont {Ponte}, \citenamefont {Qiu}, \citenamefont {Rascle}, \citenamefont {Ubelmann}, \citenamefont {Wang},\ and\ \citenamefont {Zaron}}]{Morrow_etal_2019}%
  \BibitemOpen
  \bibfield  {author} {\bibinfo {author} {\bibfnamefont {R.}~\bibnamefont {Morrow}}, \bibinfo {author} {\bibfnamefont {L.-L.}\ \bibnamefont {Fu}}, \bibinfo {author} {\bibfnamefont {F.}~\bibnamefont {Ardhuin}}, \bibinfo {author} {\bibfnamefont {M.}~\bibnamefont {Benkiran}}, \bibinfo {author} {\bibfnamefont {B.}~\bibnamefont {Chapron}}, \bibinfo {author} {\bibfnamefont {E.}~\bibnamefont {Cosme}}, \bibinfo {author} {\bibfnamefont {F.}~\bibnamefont {{d'Ovidio}}}, \bibinfo {author} {\bibfnamefont {J.~T.}\ \bibnamefont {Farrar}}, \bibinfo {author} {\bibfnamefont {S.~T.}\ \bibnamefont {Gille}}, \bibinfo {author} {\bibfnamefont {G.}~\bibnamefont {Lapeyre}}, \bibinfo {author} {\bibfnamefont {P.-Y.}\ \bibnamefont {{Le Traon}}}, \bibinfo {author} {\bibfnamefont {A.}~\bibnamefont {Pascual}}, \bibinfo {author} {\bibfnamefont {A.}~\bibnamefont {Ponte}}, \bibinfo {author} {\bibfnamefont {B.}~\bibnamefont {Qiu}}, \bibinfo {author} {\bibfnamefont {N.}~\bibnamefont {Rascle}}, \bibinfo {author} {\bibfnamefont {C.}~\bibnamefont
  {Ubelmann}}, \bibinfo {author} {\bibfnamefont {J.}~\bibnamefont {Wang}},\ and\ \bibinfo {author} {\bibfnamefont {E.~D.}\ \bibnamefont {Zaron}},\ }\bibfield  {title} {\bibinfo {title} {Global observations of fine-scale ocean surface topography with the surface water and ocean topography ({SWOT}) mission},\ }\href@noop {} {\bibfield  {journal} {\bibinfo  {journal} {Front. Mar. Sci.}\ }\textbf {\bibinfo {volume} {6}},\ \bibinfo {pages} {232} (\bibinfo {year} {2019})}\BibitemShut {NoStop}%
\bibitem [{\citenamefont {Barcel{\'o}-Llull}\ \emph {et~al.}(2021)\citenamefont {Barcel{\'o}-Llull}, \citenamefont {Pascual}, \citenamefont {S{\'a}nchez-Rom{\'a}n}, \citenamefont {Cutolo}, \citenamefont {d'Ovidio}, \citenamefont {Fifani}, \citenamefont {Ser-Giacomi}, \citenamefont {Ruiz}, \citenamefont {Mason}, \citenamefont {Cyr}, \citenamefont {Doglioli}, \citenamefont {Mourre}, \citenamefont {Allen}, \citenamefont {Alou-Font}, \citenamefont {Casas}, \citenamefont {D\'iaz-Barroso}, \citenamefont {Dumas}, \citenamefont {G\'omez-Navarro},\ and\ \citenamefont {Mu{\~{n}}oz}}]{barcelo2021fine}%
  \BibitemOpen
  \bibfield  {author} {\bibinfo {author} {\bibfnamefont {B.}~\bibnamefont {Barcel{\'o}-Llull}}, \bibinfo {author} {\bibfnamefont {A.}~\bibnamefont {Pascual}}, \bibinfo {author} {\bibfnamefont {A.}~\bibnamefont {S{\'a}nchez-Rom{\'a}n}}, \bibinfo {author} {\bibfnamefont {E.}~\bibnamefont {Cutolo}}, \bibinfo {author} {\bibfnamefont {F.}~\bibnamefont {d'Ovidio}}, \bibinfo {author} {\bibfnamefont {G.}~\bibnamefont {Fifani}}, \bibinfo {author} {\bibfnamefont {E.}~\bibnamefont {Ser-Giacomi}}, \bibinfo {author} {\bibfnamefont {S.}~\bibnamefont {Ruiz}}, \bibinfo {author} {\bibfnamefont {E.}~\bibnamefont {Mason}}, \bibinfo {author} {\bibfnamefont {F.}~\bibnamefont {Cyr}}, \bibinfo {author} {\bibfnamefont {A.}~\bibnamefont {Doglioli}}, \bibinfo {author} {\bibfnamefont {B.}~\bibnamefont {Mourre}}, \bibinfo {author} {\bibfnamefont {J.~T.}\ \bibnamefont {Allen}}, \bibinfo {author} {\bibfnamefont {E.}~\bibnamefont {Alou-Font}}, \bibinfo {author} {\bibfnamefont {B.}~\bibnamefont {Casas}}, \bibinfo {author} {\bibfnamefont
  {L.}~\bibnamefont {D\'iaz-Barroso}}, \bibinfo {author} {\bibfnamefont {F.}~\bibnamefont {Dumas}}, \bibinfo {author} {\bibfnamefont {L.}~\bibnamefont {G\'omez-Navarro}},\ and\ \bibinfo {author} {\bibfnamefont {C.}~\bibnamefont {Mu{\~{n}}oz}},\ }\bibfield  {title} {\bibinfo {title} {Fine-scale ocean currents derived from {\it in situ} observations in anticipation of the upcoming {SWOT} altimetric mission},\ }\href@noop {} {\bibfield  {journal} {\bibinfo  {journal} {Front. Mar. Sci.}\ }\textbf {\bibinfo {volume} {8}},\ \bibinfo {pages} {679844} (\bibinfo {year} {2021})}\BibitemShut {NoStop}%
\bibitem [{\citenamefont {Su}\ \emph {et~al.}(2018)\citenamefont {Su}, \citenamefont {Wang}, \citenamefont {Klein}, \citenamefont {Thompson},\ and\ \citenamefont {Menemenlis}}]{su18}%
  \BibitemOpen
  \bibfield  {author} {\bibinfo {author} {\bibfnamefont {Z.}~\bibnamefont {Su}}, \bibinfo {author} {\bibfnamefont {J.~B.}\ \bibnamefont {Wang}}, \bibinfo {author} {\bibfnamefont {P.}~\bibnamefont {Klein}}, \bibinfo {author} {\bibfnamefont {A.~F.}\ \bibnamefont {Thompson}},\ and\ \bibinfo {author} {\bibfnamefont {D.}~\bibnamefont {Menemenlis}},\ }\bibfield  {title} {\bibinfo {title} {Ocean submesoscales as a key component of the global heat budget},\ }\href@noop {} {\bibfield  {journal} {\bibinfo  {journal} {Nature Comm.}\ }\textbf {\bibinfo {volume} {9}},\ \bibinfo {pages} {775} (\bibinfo {year} {2018})}\BibitemShut {NoStop}%
\bibitem [{\citenamefont {Lumpkin}\ and\ \citenamefont {Elipot}(2010)}]{LE2010}%
  \BibitemOpen
  \bibfield  {author} {\bibinfo {author} {\bibfnamefont {R.}~\bibnamefont {Lumpkin}}\ and\ \bibinfo {author} {\bibfnamefont {S.}~\bibnamefont {Elipot}},\ }\bibfield  {title} {\bibinfo {title} {Surface drifter pair spreading in the {North Atlantic}},\ }\href@noop {} {\bibfield  {journal} {\bibinfo  {journal} {J. Geophys. Res.}\ }\textbf {\bibinfo {volume} {115}},\ \bibinfo {pages} {C12017} (\bibinfo {year} {2010})}\BibitemShut {NoStop}%
\bibitem [{\citenamefont {Berti}\ \emph {et~al.}(2011)\citenamefont {Berti}, \citenamefont {{Dos Santos}}, \citenamefont {Lacorata},\ and\ \citenamefont {Vulpiani}}]{BDLV2011}%
  \BibitemOpen
  \bibfield  {author} {\bibinfo {author} {\bibfnamefont {S.}~\bibnamefont {Berti}}, \bibinfo {author} {\bibfnamefont {F.}~\bibnamefont {{Dos Santos}}}, \bibinfo {author} {\bibfnamefont {G.}~\bibnamefont {Lacorata}},\ and\ \bibinfo {author} {\bibfnamefont {A.}~\bibnamefont {Vulpiani}},\ }\bibfield  {title} {\bibinfo {title} {Lagrangian drifter dispersion in the {Southwestern Atlantic} ocean},\ }\href@noop {} {\bibfield  {journal} {\bibinfo  {journal} {J. Phys. Oceanogr.}\ }\textbf {\bibinfo {volume} {41}},\ \bibinfo {pages} {1659} (\bibinfo {year} {2011})}\BibitemShut {NoStop}%
\bibitem [{\citenamefont {Poje}\ \emph {et~al.}(2014)\citenamefont {Poje}, \citenamefont {\"Ozg\"okmen}, \citenamefont {{Lipphardt Jr.}}, \citenamefont {Haus}, \citenamefont {Ryan}, \citenamefont {Haza}, \citenamefont {Jacobs}, \citenamefont {Reniers}, \citenamefont {Olascoaga}, \citenamefont {Novelli}, \citenamefont {Griffa}, \citenamefont {Beron-Vera}, \citenamefont {Chen}, \citenamefont {Coelho}, \citenamefont {Hogan}, \citenamefont {{Kirwan Jr.}}, \citenamefont {Huntley},\ and\ \citenamefont {Mariano}}]{Poje_etal_2014}%
  \BibitemOpen
  \bibfield  {author} {\bibinfo {author} {\bibfnamefont {A.~C.}\ \bibnamefont {Poje}}, \bibinfo {author} {\bibfnamefont {T.~M.}\ \bibnamefont {\"Ozg\"okmen}}, \bibinfo {author} {\bibfnamefont {B.~L.}\ \bibnamefont {{Lipphardt Jr.}}}, \bibinfo {author} {\bibfnamefont {B.~K.}\ \bibnamefont {Haus}}, \bibinfo {author} {\bibfnamefont {E.~H.}\ \bibnamefont {Ryan}}, \bibinfo {author} {\bibfnamefont {A.~C.}\ \bibnamefont {Haza}}, \bibinfo {author} {\bibfnamefont {G.~A.}\ \bibnamefont {Jacobs}}, \bibinfo {author} {\bibfnamefont {A.~J. H.~M.}\ \bibnamefont {Reniers}}, \bibinfo {author} {\bibfnamefont {M.~J.}\ \bibnamefont {Olascoaga}}, \bibinfo {author} {\bibfnamefont {G.}~\bibnamefont {Novelli}}, \bibinfo {author} {\bibfnamefont {A.}~\bibnamefont {Griffa}}, \bibinfo {author} {\bibfnamefont {F.~J.}\ \bibnamefont {Beron-Vera}}, \bibinfo {author} {\bibfnamefont {S.~S.}\ \bibnamefont {Chen}}, \bibinfo {author} {\bibfnamefont {E.}~\bibnamefont {Coelho}}, \bibinfo {author} {\bibfnamefont {P.~J.}\ \bibnamefont {Hogan}},
  \bibinfo {author} {\bibfnamefont {A.~D.}\ \bibnamefont {{Kirwan Jr.}}}, \bibinfo {author} {\bibfnamefont {H.~S.}\ \bibnamefont {Huntley}},\ and\ \bibinfo {author} {\bibfnamefont {A.~J.}\ \bibnamefont {Mariano}},\ }\bibfield  {title} {\bibinfo {title} {Submesoscale dispersion in the vicinity of the {Deepwater Horizon Spill}},\ }\href@noop {} {\bibfield  {journal} {\bibinfo  {journal} {Proc. Natl. Acad. Sci. U. S. A.}\ }\textbf {\bibinfo {volume} {111}},\ \bibinfo {pages} {12693} (\bibinfo {year} {2014})}\BibitemShut {NoStop}%
\bibitem [{\citenamefont {Corrado}\ \emph {et~al.}(2017)\citenamefont {Corrado}, \citenamefont {Lacorata}, \citenamefont {Palatella}, \citenamefont {Santoleri},\ and\ \citenamefont {Zambianchi}}]{CLPSZ2017}%
  \BibitemOpen
  \bibfield  {author} {\bibinfo {author} {\bibfnamefont {R.}~\bibnamefont {Corrado}}, \bibinfo {author} {\bibfnamefont {G.}~\bibnamefont {Lacorata}}, \bibinfo {author} {\bibfnamefont {L.}~\bibnamefont {Palatella}}, \bibinfo {author} {\bibfnamefont {R.}~\bibnamefont {Santoleri}},\ and\ \bibinfo {author} {\bibfnamefont {E.}~\bibnamefont {Zambianchi}},\ }\bibfield  {title} {\bibinfo {title} {General characteristics of relative dispersion in the ocean},\ }\href@noop {} {\bibfield  {journal} {\bibinfo  {journal} {Sci. Rep.}\ }\textbf {\bibinfo {volume} {7}},\ \bibinfo {pages} {46291} (\bibinfo {year} {2017})}\BibitemShut {NoStop}%
\bibitem [{\citenamefont {D'Asaro}\ \emph {et~al.}(2018)\citenamefont {D'Asaro}, \citenamefont {Shcherbina}, \citenamefont {Klymak}, \citenamefont {Molemaker}, \citenamefont {Novelli}, \citenamefont {Guigand}, \citenamefont {Haza}, \citenamefont {Haus}, \citenamefont {Ryan}, \citenamefont {Jacobs}, \citenamefont {Huntley}, \citenamefont {Laxague}, \citenamefont {Chen}, \citenamefont {Judt}, \citenamefont {McWilliams}, \citenamefont {Barkan}, \citenamefont {{Kirwan Jr.}}, \citenamefont {Poje},\ and\ \citenamefont {\"Ozg\"okmen}}]{Dasaro_etal_2018}%
  \BibitemOpen
  \bibfield  {author} {\bibinfo {author} {\bibfnamefont {E.~A.}\ \bibnamefont {D'Asaro}}, \bibinfo {author} {\bibfnamefont {A.~Y.}\ \bibnamefont {Shcherbina}}, \bibinfo {author} {\bibfnamefont {J.~M.}\ \bibnamefont {Klymak}}, \bibinfo {author} {\bibfnamefont {J.}~\bibnamefont {Molemaker}}, \bibinfo {author} {\bibfnamefont {G.}~\bibnamefont {Novelli}}, \bibinfo {author} {\bibfnamefont {C.~M.}\ \bibnamefont {Guigand}}, \bibinfo {author} {\bibfnamefont {A.~C.}\ \bibnamefont {Haza}}, \bibinfo {author} {\bibfnamefont {B.~K.}\ \bibnamefont {Haus}}, \bibinfo {author} {\bibfnamefont {E.~H.}\ \bibnamefont {Ryan}}, \bibinfo {author} {\bibfnamefont {G.~A.}\ \bibnamefont {Jacobs}}, \bibinfo {author} {\bibfnamefont {H.~S.}\ \bibnamefont {Huntley}}, \bibinfo {author} {\bibfnamefont {N.~J.~M.}\ \bibnamefont {Laxague}}, \bibinfo {author} {\bibfnamefont {S.}~\bibnamefont {Chen}}, \bibinfo {author} {\bibfnamefont {F.}~\bibnamefont {Judt}}, \bibinfo {author} {\bibfnamefont {J.~C.}\ \bibnamefont {McWilliams}}, \bibinfo {author}
  {\bibfnamefont {R.}~\bibnamefont {Barkan}}, \bibinfo {author} {\bibfnamefont {A.~D.}\ \bibnamefont {{Kirwan Jr.}}}, \bibinfo {author} {\bibfnamefont {A.~C.}\ \bibnamefont {Poje}},\ and\ \bibinfo {author} {\bibfnamefont {T.~M.}\ \bibnamefont {\"Ozg\"okmen}},\ }\bibfield  {title} {\bibinfo {title} {Ocean convergence and the dispersion of flotsam},\ }\href@noop {} {\bibfield  {journal} {\bibinfo  {journal} {Proc. Natl. Acad. Sci. U. S. A.}\ }\textbf {\bibinfo {volume} {115}},\ \bibinfo {pages} {1162} (\bibinfo {year} {2018})}\BibitemShut {NoStop}%
\bibitem [{\citenamefont {Vic}\ \emph {et~al.}(2022)\citenamefont {Vic}, \citenamefont {{Hasco\"et}}, \citenamefont {Gula}, \citenamefont {Huck},\ and\ \citenamefont {Maes}}]{vic2022}%
  \BibitemOpen
  \bibfield  {author} {\bibinfo {author} {\bibfnamefont {C.}~\bibnamefont {Vic}}, \bibinfo {author} {\bibfnamefont {S.}~\bibnamefont {{Hasco\"et}}}, \bibinfo {author} {\bibfnamefont {J.}~\bibnamefont {Gula}}, \bibinfo {author} {\bibfnamefont {T.}~\bibnamefont {Huck}},\ and\ \bibinfo {author} {\bibfnamefont {C.}~\bibnamefont {Maes}},\ }\bibfield  {title} {\bibinfo {title} {Oceanic mesoscale cyclones cluster surface {Lagrangian} material},\ }\href@noop {} {\bibfield  {journal} {\bibinfo  {journal} {Geophys. Res. Lett.}\ }\textbf {\bibinfo {volume} {49}},\ \bibinfo {pages} {e2021GL097488} (\bibinfo {year} {2022})}\BibitemShut {NoStop}%
\bibitem [{\citenamefont {Balwada}\ \emph {et~al.}(2022)\citenamefont {Balwada}, \citenamefont {Xie}, \citenamefont {Marino},\ and\ \citenamefont {Feraco}}]{balwada2022direct}%
  \BibitemOpen
  \bibfield  {author} {\bibinfo {author} {\bibfnamefont {D.}~\bibnamefont {Balwada}}, \bibinfo {author} {\bibfnamefont {J.-H.}\ \bibnamefont {Xie}}, \bibinfo {author} {\bibfnamefont {R.}~\bibnamefont {Marino}},\ and\ \bibinfo {author} {\bibfnamefont {F.}~\bibnamefont {Feraco}},\ }\bibfield  {title} {\bibinfo {title} {Direct observational evidence of an oceanic dual kinetic energy cascade and its seasonality},\ }\href@noop {} {\bibfield  {journal} {\bibinfo  {journal} {Sci. Adv.}\ }\textbf {\bibinfo {volume} {8}},\ \bibinfo {pages} {eabq2566} (\bibinfo {year} {2022})}\BibitemShut {NoStop}%
\bibitem [{\citenamefont {Fu}\ \emph {et~al.}(2010)\citenamefont {Fu}, \citenamefont {Chelton}, \citenamefont {{Le Traon}},\ and\ \citenamefont {Morrow}}]{fu2010eddy}%
  \BibitemOpen
  \bibfield  {author} {\bibinfo {author} {\bibfnamefont {L.-L.}\ \bibnamefont {Fu}}, \bibinfo {author} {\bibfnamefont {D.~B.}\ \bibnamefont {Chelton}}, \bibinfo {author} {\bibfnamefont {P.-Y.}\ \bibnamefont {{Le Traon}}},\ and\ \bibinfo {author} {\bibfnamefont {R.}~\bibnamefont {Morrow}},\ }\bibfield  {title} {\bibinfo {title} {Eddy dynamics from satellite altimetry},\ }\href@noop {} {\bibfield  {journal} {\bibinfo  {journal} {Oceanography}\ }\textbf {\bibinfo {volume} {23}},\ \bibinfo {pages} {14} (\bibinfo {year} {2010})}\BibitemShut {NoStop}%
\bibitem [{\citenamefont {Morrow}\ \emph {et~al.}(2023)\citenamefont {Morrow}, \citenamefont {Fu}, \citenamefont {Rio}, \citenamefont {R.~Ray}, \citenamefont {Prandi}, \citenamefont {{Le Traon}},\ and\ \citenamefont {Benveniste}}]{morrow2023ocean}%
  \BibitemOpen
  \bibfield  {author} {\bibinfo {author} {\bibfnamefont {R.}~\bibnamefont {Morrow}}, \bibinfo {author} {\bibfnamefont {L.-L.}\ \bibnamefont {Fu}}, \bibinfo {author} {\bibfnamefont {M.-H.}\ \bibnamefont {Rio}}, \bibinfo {author} {\bibfnamefont {R.}~\bibnamefont {R.~Ray}}, \bibinfo {author} {\bibfnamefont {P.}~\bibnamefont {Prandi}}, \bibinfo {author} {\bibfnamefont {P.-Y.}\ \bibnamefont {{Le Traon}}},\ and\ \bibinfo {author} {\bibfnamefont {J.}~\bibnamefont {Benveniste}},\ }\bibfield  {title} {\bibinfo {title} {Ocean circulation from space},\ }\href@noop {} {\bibfield  {journal} {\bibinfo  {journal} {Surv. Geophys.}\ }\textbf {\bibinfo {volume} {44}},\ \bibinfo {pages} {1243} (\bibinfo {year} {2023})}\BibitemShut {NoStop}%
\bibitem [{\citenamefont {Fu}\ \emph {et~al.}(2024)\citenamefont {Fu}, \citenamefont {Pavelsky}, \citenamefont {Cretaux}, \citenamefont {Morrow}, \citenamefont {Farrar}, \citenamefont {Vaze}, \citenamefont {Sengenes}, \citenamefont {Vinogradova-Shiffer}, \citenamefont {Sylvestre-Baron}, \citenamefont {Picot},\ and\ \citenamefont {Dibarboure}}]{fu2024surface}%
  \BibitemOpen
  \bibfield  {author} {\bibinfo {author} {\bibfnamefont {L.-L.}\ \bibnamefont {Fu}}, \bibinfo {author} {\bibfnamefont {T.}~\bibnamefont {Pavelsky}}, \bibinfo {author} {\bibfnamefont {J.-F.}\ \bibnamefont {Cretaux}}, \bibinfo {author} {\bibfnamefont {R.}~\bibnamefont {Morrow}}, \bibinfo {author} {\bibfnamefont {J.~T.}\ \bibnamefont {Farrar}}, \bibinfo {author} {\bibfnamefont {P.}~\bibnamefont {Vaze}}, \bibinfo {author} {\bibfnamefont {P.}~\bibnamefont {Sengenes}}, \bibinfo {author} {\bibfnamefont {N.}~\bibnamefont {Vinogradova-Shiffer}}, \bibinfo {author} {\bibfnamefont {A.}~\bibnamefont {Sylvestre-Baron}}, \bibinfo {author} {\bibfnamefont {N.}~\bibnamefont {Picot}},\ and\ \bibinfo {author} {\bibfnamefont {G.}~\bibnamefont {Dibarboure}},\ }\bibfield  {title} {\bibinfo {title} {The surface water and ocean topography mission: A breakthrough in radar remote sensing of the ocean and land surface water},\ }\href@noop {} {\bibfield  {journal} {\bibinfo  {journal} {Geophy. Res. Lett.}\ }\textbf {\bibinfo {volume}
  {51}},\ \bibinfo {pages} {e2023GL107652} (\bibinfo {year} {2024})}\BibitemShut {NoStop}%
\bibitem [{\citenamefont {Rocha}\ \emph {et~al.}(2016)\citenamefont {Rocha}, \citenamefont {Chereskin}, \citenamefont {Gille},\ and\ \citenamefont {Menemenlis}}]{Rocha2016}%
  \BibitemOpen
  \bibfield  {author} {\bibinfo {author} {\bibfnamefont {C.~B.}\ \bibnamefont {Rocha}}, \bibinfo {author} {\bibfnamefont {T.~K.}\ \bibnamefont {Chereskin}}, \bibinfo {author} {\bibfnamefont {S.~T.}\ \bibnamefont {Gille}},\ and\ \bibinfo {author} {\bibfnamefont {D.}~\bibnamefont {Menemenlis}},\ }\bibfield  {title} {\bibinfo {title} {Mesoscale to submesoscale wavenumber spectra in {Drake} passage},\ }\href@noop {} {\bibfield  {journal} {\bibinfo  {journal} {J. Phys. Oceanog.}\ }\textbf {\bibinfo {volume} {46}},\ \bibinfo {pages} {601} (\bibinfo {year} {2016})}\BibitemShut {NoStop}%
\bibitem [{\citenamefont {Wang}\ \emph {et~al.}(2019)\citenamefont {Wang}, \citenamefont {Fu}, \citenamefont {Torres}, \citenamefont {Chen}, \citenamefont {Qiu},\ and\ \citenamefont {Menemenlis}}]{Wang2019}%
  \BibitemOpen
  \bibfield  {author} {\bibinfo {author} {\bibfnamefont {J.}~\bibnamefont {Wang}}, \bibinfo {author} {\bibfnamefont {L.}~\bibnamefont {Fu}}, \bibinfo {author} {\bibfnamefont {H.~S.}\ \bibnamefont {Torres}}, \bibinfo {author} {\bibfnamefont {S.}~\bibnamefont {Chen}}, \bibinfo {author} {\bibfnamefont {B.}~\bibnamefont {Qiu}},\ and\ \bibinfo {author} {\bibfnamefont {D.}~\bibnamefont {Menemenlis}},\ }\bibfield  {title} {\bibinfo {title} {On the spatial scales to be resolved by the surface water and ocean topography ka-band radar interferometer},\ }\href@noop {} {\bibfield  {journal} {\bibinfo  {journal} {J. Atmos. Oceanic Technol.}\ }\textbf {\bibinfo {volume} {36}},\ \bibinfo {pages} {87} (\bibinfo {year} {2019})}\BibitemShut {NoStop}%
\bibitem [{\citenamefont {Callies}\ \emph {et~al.}(2016)\citenamefont {Callies}, \citenamefont {Flierl}, \citenamefont {Ferrari},\ and\ \citenamefont {Fox-Kemper}}]{Callies_etal2016}%
  \BibitemOpen
  \bibfield  {author} {\bibinfo {author} {\bibfnamefont {J.}~\bibnamefont {Callies}}, \bibinfo {author} {\bibfnamefont {G.}~\bibnamefont {Flierl}}, \bibinfo {author} {\bibfnamefont {R.}~\bibnamefont {Ferrari}},\ and\ \bibinfo {author} {\bibfnamefont {B.}~\bibnamefont {Fox-Kemper}},\ }\bibfield  {title} {\bibinfo {title} {The role of mixed-layer instabilities in submesoscale turbulence},\ }\href@noop {} {\bibfield  {journal} {\bibinfo  {journal} {J. Fluid Mech.}\ }\textbf {\bibinfo {volume} {788}},\ \bibinfo {pages} {5} (\bibinfo {year} {2016})}\BibitemShut {NoStop}%
\bibitem [{\citenamefont {Berti}\ and\ \citenamefont {Lapeyre}(2021)}]{BL2021}%
  \BibitemOpen
  \bibfield  {author} {\bibinfo {author} {\bibfnamefont {S.}~\bibnamefont {Berti}}\ and\ \bibinfo {author} {\bibfnamefont {G.}~\bibnamefont {Lapeyre}},\ }\bibfield  {title} {\bibinfo {title} {Lagrangian pair dispersion in upper-ocean turbulence in the presence of mixed-layer instabilities},\ }\href@noop {} {\bibfield  {journal} {\bibinfo  {journal} {Phys. Fluids}\ }\textbf {\bibinfo {volume} {33}},\ \bibinfo {pages} {036603} (\bibinfo {year} {2021})}\BibitemShut {NoStop}%
\bibitem [{\citenamefont {Lapeyre}\ and\ \citenamefont {Klein}(2006)}]{lapeyre2006}%
  \BibitemOpen
  \bibfield  {author} {\bibinfo {author} {\bibfnamefont {G.}~\bibnamefont {Lapeyre}}\ and\ \bibinfo {author} {\bibfnamefont {P.}~\bibnamefont {Klein}},\ }\bibfield  {title} {\bibinfo {title} {Dynamics of the upper oceanic layers in terms of surface quasigeostrophy theory},\ }\href@noop {} {\bibfield  {journal} {\bibinfo  {journal} {J. Phys. Oceanog.}\ }\textbf {\bibinfo {volume} {36}},\ \bibinfo {pages} {165} (\bibinfo {year} {2006})}\BibitemShut {NoStop}%
\bibitem [{\citenamefont {Held}\ \emph {et~al.}(1995)\citenamefont {Held}, \citenamefont {Pierrehumbert}, \citenamefont {Garner},\ and\ \citenamefont {Swanson}}]{HPGS1995}%
  \BibitemOpen
  \bibfield  {author} {\bibinfo {author} {\bibfnamefont {I.~M.}\ \bibnamefont {Held}}, \bibinfo {author} {\bibfnamefont {R.~T.}\ \bibnamefont {Pierrehumbert}}, \bibinfo {author} {\bibfnamefont {S.~T.}\ \bibnamefont {Garner}},\ and\ \bibinfo {author} {\bibfnamefont {K.~L.}\ \bibnamefont {Swanson}},\ }\bibfield  {title} {\bibinfo {title} {Surface quasi-geostrophic dynamics},\ }\href@noop {} {\bibfield  {journal} {\bibinfo  {journal} {J. Fluid Mech.}\ }\textbf {\bibinfo {volume} {282}},\ \bibinfo {pages} {1} (\bibinfo {year} {1995})}\BibitemShut {NoStop}%
\bibitem [{\citenamefont {Lapeyre}(2017)}]{lapeyre2017}%
  \BibitemOpen
  \bibfield  {author} {\bibinfo {author} {\bibfnamefont {G.}~\bibnamefont {Lapeyre}},\ }\bibfield  {title} {\bibinfo {title} {Surface quasi-geostrophy},\ }\href@noop {} {\bibfield  {journal} {\bibinfo  {journal} {Fluids}\ }\textbf {\bibinfo {volume} {2}} (\bibinfo {year} {2017})}\BibitemShut {NoStop}%
\bibitem [{\citenamefont {Rudnick}(2001)}]{Rudnick2001}%
  \BibitemOpen
  \bibfield  {author} {\bibinfo {author} {\bibfnamefont {D.~L.}\ \bibnamefont {Rudnick}},\ }\bibfield  {title} {\bibinfo {title} {On the skewness of vorticity in the upper ocean},\ }\href@noop {} {\bibfield  {journal} {\bibinfo  {journal} {Geophys. Res. Lett.}\ }\textbf {\bibinfo {volume} {28}},\ \bibinfo {pages} {2045} (\bibinfo {year} {2001})}\BibitemShut {NoStop}%
\bibitem [{\citenamefont {Roullet}\ and\ \citenamefont {Klein}(2010)}]{RK2010}%
  \BibitemOpen
  \bibfield  {author} {\bibinfo {author} {\bibfnamefont {G.}~\bibnamefont {Roullet}}\ and\ \bibinfo {author} {\bibfnamefont {P.}~\bibnamefont {Klein}},\ }\bibfield  {title} {\bibinfo {title} {Cyclone-anticyclone asymmetry in geophysical turbulence},\ }\href@noop {} {\bibfield  {journal} {\bibinfo  {journal} {Phys. Rev. Lett.}\ }\textbf {\bibinfo {volume} {104}},\ \bibinfo {pages} {218501} (\bibinfo {year} {2010})}\BibitemShut {NoStop}%
\bibitem [{\citenamefont {Shcherbina}\ \emph {et~al.}(2013)\citenamefont {Shcherbina}, \citenamefont {D'Asaro}, \citenamefont {Lee}, \citenamefont {Klymak}, \citenamefont {Molemaker},\ and\ \citenamefont {McWilliams}}]{Shcherbina_etal_2013}%
  \BibitemOpen
  \bibfield  {author} {\bibinfo {author} {\bibfnamefont {A.~Y.}\ \bibnamefont {Shcherbina}}, \bibinfo {author} {\bibfnamefont {E.~A.}\ \bibnamefont {D'Asaro}}, \bibinfo {author} {\bibfnamefont {C.~M.}\ \bibnamefont {Lee}}, \bibinfo {author} {\bibfnamefont {J.~M.}\ \bibnamefont {Klymak}}, \bibinfo {author} {\bibfnamefont {M.~J.}\ \bibnamefont {Molemaker}},\ and\ \bibinfo {author} {\bibfnamefont {J.~C.}\ \bibnamefont {McWilliams}},\ }\bibfield  {title} {\bibinfo {title} {Statistics of vertical vorticity, divergence, and strain in a developed submesoscale turbulence field},\ }\href@noop {} {\bibfield  {journal} {\bibinfo  {journal} {Geophys. Res. Lett.}\ }\textbf {\bibinfo {volume} {40}},\ \bibinfo {pages} {4706} (\bibinfo {year} {2013})}\BibitemShut {NoStop}%
\bibitem [{\citenamefont {Buckingham}\ \emph {et~al.}(2016)\citenamefont {Buckingham}, \citenamefont {Garabato}, \citenamefont {Thompson}, \citenamefont {Brannigan}, \citenamefont {Lazar}, \citenamefont {Marshall}, \citenamefont {Nurser}, \citenamefont {Damerell}, \citenamefont {Heywood},\ and\ \citenamefont {Belcher}}]{buckingham2016}%
  \BibitemOpen
  \bibfield  {author} {\bibinfo {author} {\bibfnamefont {C.~E.}\ \bibnamefont {Buckingham}}, \bibinfo {author} {\bibfnamefont {A.~C.~N.}\ \bibnamefont {Garabato}}, \bibinfo {author} {\bibfnamefont {A.~F.}\ \bibnamefont {Thompson}}, \bibinfo {author} {\bibfnamefont {L.}~\bibnamefont {Brannigan}}, \bibinfo {author} {\bibfnamefont {A.}~\bibnamefont {Lazar}}, \bibinfo {author} {\bibfnamefont {D.~P.}\ \bibnamefont {Marshall}}, \bibinfo {author} {\bibfnamefont {A.~J.~G.}\ \bibnamefont {Nurser}}, \bibinfo {author} {\bibfnamefont {G.}~\bibnamefont {Damerell}}, \bibinfo {author} {\bibfnamefont {K.~J.}\ \bibnamefont {Heywood}},\ and\ \bibinfo {author} {\bibfnamefont {S.~E.}\ \bibnamefont {Belcher}},\ }\bibfield  {title} {\bibinfo {title} {Seasonality of submesoscale flows in the ocean surface boundary layer},\ }\href@noop {} {\bibfield  {journal} {\bibinfo  {journal} {Geophys. Res. Lett.}\ }\textbf {\bibinfo {volume} {43}},\ \bibinfo {pages} {2118} (\bibinfo {year} {2016})}\BibitemShut {NoStop}%
\bibitem [{\citenamefont {Jacobs}\ \emph {et~al.}(2016)\citenamefont {Jacobs}, \citenamefont {Huntley}, \citenamefont {{Kirwan Jr.}}, \citenamefont {{Lipphardt Jr.}}, \citenamefont {Campbell}, \citenamefont {Smith}, \citenamefont {Edwards},\ and\ \citenamefont {Bartels}}]{Jacobs_etal_2016}%
  \BibitemOpen
  \bibfield  {author} {\bibinfo {author} {\bibfnamefont {G.~A.}\ \bibnamefont {Jacobs}}, \bibinfo {author} {\bibfnamefont {H.~S.}\ \bibnamefont {Huntley}}, \bibinfo {author} {\bibfnamefont {A.~D.}\ \bibnamefont {{Kirwan Jr.}}}, \bibinfo {author} {\bibfnamefont {B.~L.}\ \bibnamefont {{Lipphardt Jr.}}}, \bibinfo {author} {\bibfnamefont {T.}~\bibnamefont {Campbell}}, \bibinfo {author} {\bibfnamefont {T.}~\bibnamefont {Smith}}, \bibinfo {author} {\bibfnamefont {K.}~\bibnamefont {Edwards}},\ and\ \bibinfo {author} {\bibfnamefont {B.}~\bibnamefont {Bartels}},\ }\bibfield  {title} {\bibinfo {title} {Ocean processes underlying surface clustering},\ }\href@noop {} {\bibfield  {journal} {\bibinfo  {journal} {J. Geophys. Res.}\ }\textbf {\bibinfo {volume} {121}},\ \bibinfo {pages} {180} (\bibinfo {year} {2016})}\BibitemShut {NoStop}%
\bibitem [{\citenamefont {Berta}\ \emph {et~al.}(2020)\citenamefont {Berta}, \citenamefont {Griffa}, \citenamefont {Haza}, \citenamefont {Horstmann}, \citenamefont {Huntley}, \citenamefont {Ibrahim}, \citenamefont {Lund}, \citenamefont {\"Ozg\"okmen},\ and\ \citenamefont {Poje}}]{Berta_etal_2020}%
  \BibitemOpen
  \bibfield  {author} {\bibinfo {author} {\bibfnamefont {M.}~\bibnamefont {Berta}}, \bibinfo {author} {\bibfnamefont {A.}~\bibnamefont {Griffa}}, \bibinfo {author} {\bibfnamefont {A.~C.}\ \bibnamefont {Haza}}, \bibinfo {author} {\bibfnamefont {J.}~\bibnamefont {Horstmann}}, \bibinfo {author} {\bibfnamefont {H.~S.}\ \bibnamefont {Huntley}}, \bibinfo {author} {\bibfnamefont {R.}~\bibnamefont {Ibrahim}}, \bibinfo {author} {\bibfnamefont {B.}~\bibnamefont {Lund}}, \bibinfo {author} {\bibfnamefont {T.~M.}\ \bibnamefont {\"Ozg\"okmen}},\ and\ \bibinfo {author} {\bibfnamefont {A.~C.}\ \bibnamefont {Poje}},\ }\bibfield  {title} {\bibinfo {title} {Submesoscale kinematic properties in summer and winter surface flows in the {Northern Gulf of Mexico}},\ }\href@noop {} {\bibfield  {journal} {\bibinfo  {journal} {J. Geophys. Res.}\ }\textbf {\bibinfo {volume} {125}},\ \bibinfo {pages} {e2020JC016085} (\bibinfo {year} {2020})}\BibitemShut {NoStop}%
\bibitem [{\citenamefont {Ragone}\ and\ \citenamefont {Badin}(2016)}]{ragone_badin_2016}%
  \BibitemOpen
  \bibfield  {author} {\bibinfo {author} {\bibfnamefont {F.}~\bibnamefont {Ragone}}\ and\ \bibinfo {author} {\bibfnamefont {G.}~\bibnamefont {Badin}},\ }\bibfield  {title} {\bibinfo {title} {A study of surface semi-geostrophic turbulence: freely decaying dynamics},\ }\href@noop {} {\bibfield  {journal} {\bibinfo  {journal} {J. Fluid Mech.}\ }\textbf {\bibinfo {volume} {792}},\ \bibinfo {pages} {740} (\bibinfo {year} {2016})}\BibitemShut {NoStop}%
\bibitem [{\citenamefont {Hakim}\ \emph {et~al.}(2002)\citenamefont {Hakim}, \citenamefont {Snyder},\ and\ \citenamefont {Muraki}}]{HSM2002}%
  \BibitemOpen
  \bibfield  {author} {\bibinfo {author} {\bibfnamefont {G.~J.}\ \bibnamefont {Hakim}}, \bibinfo {author} {\bibfnamefont {C.}~\bibnamefont {Snyder}},\ and\ \bibinfo {author} {\bibfnamefont {D.~J.}\ \bibnamefont {Muraki}},\ }\bibfield  {title} {\bibinfo {title} {A new surface model for cyclone–anticyclone asymmetry},\ }\href@noop {} {\bibfield  {journal} {\bibinfo  {journal} {J. Atmos. Sci.}\ }\textbf {\bibinfo {volume} {59}},\ \bibinfo {pages} {2405} (\bibinfo {year} {2002})}\BibitemShut {NoStop}%
\bibitem [{\citenamefont {Maalouly}\ \emph {et~al.}(2023)\citenamefont {Maalouly}, \citenamefont {Lapeyre}, \citenamefont {Cozian}, \citenamefont {Mompean},\ and\ \citenamefont {Berti}}]{maaloulyetal2023}%
  \BibitemOpen
  \bibfield  {author} {\bibinfo {author} {\bibfnamefont {M.}~\bibnamefont {Maalouly}}, \bibinfo {author} {\bibfnamefont {G.}~\bibnamefont {Lapeyre}}, \bibinfo {author} {\bibfnamefont {B.}~\bibnamefont {Cozian}}, \bibinfo {author} {\bibfnamefont {G.}~\bibnamefont {Mompean}},\ and\ \bibinfo {author} {\bibfnamefont {S.}~\bibnamefont {Berti}},\ }\bibfield  {title} {\bibinfo {title} {{Particle dispersion and clustering in surface ocean turbulence with ageostrophic dynamics}},\ }\href@noop {} {\bibfield  {journal} {\bibinfo  {journal} {Phys. Fluids}\ }\textbf {\bibinfo {volume} {35}},\ \bibinfo {pages} {126601} (\bibinfo {year} {2023})}\BibitemShut {NoStop}%
\bibitem [{\citenamefont {Foussard}\ \emph {et~al.}(2017)\citenamefont {Foussard}, \citenamefont {Berti}, \citenamefont {Perrot},\ and\ \citenamefont {Lapeyre}}]{FBPL2017}%
  \BibitemOpen
  \bibfield  {author} {\bibinfo {author} {\bibfnamefont {A.}~\bibnamefont {Foussard}}, \bibinfo {author} {\bibfnamefont {S.}~\bibnamefont {Berti}}, \bibinfo {author} {\bibfnamefont {X.}~\bibnamefont {Perrot}},\ and\ \bibinfo {author} {\bibfnamefont {G.}~\bibnamefont {Lapeyre}},\ }\bibfield  {title} {\bibinfo {title} {Relative dispersion in generalized two-dimensional turbulence},\ }\href@noop {} {\bibfield  {journal} {\bibinfo  {journal} {J. Fluid Mech.}\ }\textbf {\bibinfo {volume} {821}},\ \bibinfo {pages} {358} (\bibinfo {year} {2017})}\BibitemShut {NoStop}%
\bibitem [{\citenamefont {Perruche}\ \emph {et~al.}(2011)\citenamefont {Perruche}, \citenamefont {Rivi\`ere}, \citenamefont {Lapeyre}, \citenamefont {Carton},\ and\ \citenamefont {Pondaven}}]{Perruche_etal11}%
  \BibitemOpen
  \bibfield  {author} {\bibinfo {author} {\bibfnamefont {C.}~\bibnamefont {Perruche}}, \bibinfo {author} {\bibfnamefont {P.}~\bibnamefont {Rivi\`ere}}, \bibinfo {author} {\bibfnamefont {G.}~\bibnamefont {Lapeyre}}, \bibinfo {author} {\bibfnamefont {X.}~\bibnamefont {Carton}},\ and\ \bibinfo {author} {\bibfnamefont {P.}~\bibnamefont {Pondaven}},\ }\bibfield  {title} {\bibinfo {title} {Effects of surface quasi-geostrophic turbulence on phytoplankton competition and coexistence},\ }\href@noop {} {\bibfield  {journal} {\bibinfo  {journal} {J. Mar. Res.}\ }\textbf {\bibinfo {volume} {69}},\ \bibinfo {pages} {105} (\bibinfo {year} {2011})}\BibitemShut {NoStop}%
\bibitem [{\citenamefont {Weiss}(2022)}]{Weiss2022}%
  \BibitemOpen
  \bibfield  {author} {\bibinfo {author} {\bibfnamefont {J.~B.}\ \bibnamefont {Weiss}},\ }\bibfield  {title} {\bibinfo {title} {Point-vortex dynamics in three-dimensional ageostrophic balanced flows},\ }\href@noop {} {\bibfield  {journal} {\bibinfo  {journal} {J. Fluid Mech.}\ }\textbf {\bibinfo {volume} {936}},\ \bibinfo {pages} {A19} (\bibinfo {year} {2022})}\BibitemShut {NoStop}%
\bibitem [{\citenamefont {Yu}\ \emph {et~al.}(2019)\citenamefont {Yu}, \citenamefont {Ponte}, \citenamefont {Elipot}, \citenamefont {Menemenlis}, \citenamefont {Zaron},\ and\ \citenamefont {Abernathey}}]{yu2019surface}%
  \BibitemOpen
  \bibfield  {author} {\bibinfo {author} {\bibfnamefont {X.}~\bibnamefont {Yu}}, \bibinfo {author} {\bibfnamefont {A.~L.}\ \bibnamefont {Ponte}}, \bibinfo {author} {\bibfnamefont {S.}~\bibnamefont {Elipot}}, \bibinfo {author} {\bibfnamefont {D.}~\bibnamefont {Menemenlis}}, \bibinfo {author} {\bibfnamefont {E.~D.}\ \bibnamefont {Zaron}},\ and\ \bibinfo {author} {\bibfnamefont {R.}~\bibnamefont {Abernathey}},\ }\bibfield  {title} {\bibinfo {title} {Surface kinetic energy distributions in the global oceans from a high-resolution numerical model and surface drifter observations},\ }\href@noop {} {\bibfield  {journal} {\bibinfo  {journal} {Geophys. Res. Lett.}\ }\textbf {\bibinfo {volume} {46}},\ \bibinfo {pages} {9757} (\bibinfo {year} {2019})}\BibitemShut {NoStop}%
\bibitem [{\citenamefont {Wang}\ \emph {et~al.}(2022)\citenamefont {Wang}, \citenamefont {Grisouard}, \citenamefont {Salehipour}, \citenamefont {Nuz}, \citenamefont {Poon},\ and\ \citenamefont {Ponte}}]{wang2022deep}%
  \BibitemOpen
  \bibfield  {author} {\bibinfo {author} {\bibfnamefont {H.}~\bibnamefont {Wang}}, \bibinfo {author} {\bibfnamefont {N.}~\bibnamefont {Grisouard}}, \bibinfo {author} {\bibfnamefont {H.}~\bibnamefont {Salehipour}}, \bibinfo {author} {\bibfnamefont {A.}~\bibnamefont {Nuz}}, \bibinfo {author} {\bibfnamefont {M.}~\bibnamefont {Poon}},\ and\ \bibinfo {author} {\bibfnamefont {A.~L.}\ \bibnamefont {Ponte}},\ }\bibfield  {title} {\bibinfo {title} {A deep learning approach to extract internal tides scattered by geostrophic turbulence},\ }\href@noop {} {\bibfield  {journal} {\bibinfo  {journal} {Geophys. Res. Lett.}\ }\textbf {\bibinfo {volume} {49}},\ \bibinfo {pages} {e2022GL099400} (\bibinfo {year} {2022})}\BibitemShut {NoStop}%
\bibitem [{\citenamefont {Arbic}\ \emph {et~al.}(2022)\citenamefont {Arbic}, \citenamefont {Elipot}, \citenamefont {Brasch}, \citenamefont {Menemenlis}, \citenamefont {Ponte}, \citenamefont {Shriver}, \citenamefont {Yu}, \citenamefont {Zaron}, \citenamefont {Alford}, \citenamefont {Buijsman}, \citenamefont {Abernathey}, \citenamefont {Garcia}, \citenamefont {Guan}, \citenamefont {Martin},\ and\ \citenamefont {Nelson}}]{arbic2022near}%
  \BibitemOpen
  \bibfield  {author} {\bibinfo {author} {\bibfnamefont {B.~K.}\ \bibnamefont {Arbic}}, \bibinfo {author} {\bibfnamefont {S.}~\bibnamefont {Elipot}}, \bibinfo {author} {\bibfnamefont {J.~M.}\ \bibnamefont {Brasch}}, \bibinfo {author} {\bibfnamefont {D.}~\bibnamefont {Menemenlis}}, \bibinfo {author} {\bibfnamefont {A.~L.}\ \bibnamefont {Ponte}}, \bibinfo {author} {\bibfnamefont {J.~F.}\ \bibnamefont {Shriver}}, \bibinfo {author} {\bibfnamefont {X.}~\bibnamefont {Yu}}, \bibinfo {author} {\bibfnamefont {E.~D.}\ \bibnamefont {Zaron}}, \bibinfo {author} {\bibfnamefont {M.~H.}\ \bibnamefont {Alford}}, \bibinfo {author} {\bibfnamefont {M.~C.}\ \bibnamefont {Buijsman}}, \bibinfo {author} {\bibfnamefont {R.}~\bibnamefont {Abernathey}}, \bibinfo {author} {\bibfnamefont {D.}~\bibnamefont {Garcia}}, \bibinfo {author} {\bibfnamefont {L.}~\bibnamefont {Guan}}, \bibinfo {author} {\bibfnamefont {P.~E.}\ \bibnamefont {Martin}},\ and\ \bibinfo {author} {\bibfnamefont {A.~D.}\ \bibnamefont {Nelson}},\ }\bibfield  {title}
  {\bibinfo {title} {Near-surface oceanic kinetic energy distributions from drifter observations and numerical models},\ }\href@noop {} {\bibfield  {journal} {\bibinfo  {journal} {J. Geophys. Res.}\ }\textbf {\bibinfo {volume} {127}},\ \bibinfo {pages} {e2022JC018551} (\bibinfo {year} {2022})}\BibitemShut {NoStop}%
\bibitem [{\citenamefont {Celani}\ \emph {et~al.}(2004)\citenamefont {Celani}, \citenamefont {Cencini}, \citenamefont {Mazzino},\ and\ \citenamefont {Vergassola}}]{CCMV2004}%
  \BibitemOpen
  \bibfield  {author} {\bibinfo {author} {\bibfnamefont {A.}~\bibnamefont {Celani}}, \bibinfo {author} {\bibfnamefont {M.}~\bibnamefont {Cencini}}, \bibinfo {author} {\bibfnamefont {A.}~\bibnamefont {Mazzino}},\ and\ \bibinfo {author} {\bibfnamefont {M.}~\bibnamefont {Vergassola}},\ }\bibfield  {title} {\bibinfo {title} {Active and passive fields face to face},\ }\href@noop {} {\bibfield  {journal} {\bibinfo  {journal} {New J. Phys.}\ }\textbf {\bibinfo {volume} {6}},\ \bibinfo {pages} {72} (\bibinfo {year} {2004})}\BibitemShut {NoStop}%
\bibitem [{\citenamefont {Balwada}\ \emph {et~al.}(2021)\citenamefont {Balwada}, \citenamefont {Xiao}, \citenamefont {Smith}, \citenamefont {Abernathey},\ and\ \citenamefont {Gray}}]{Balwada_etal_2021}%
  \BibitemOpen
  \bibfield  {author} {\bibinfo {author} {\bibfnamefont {D.}~\bibnamefont {Balwada}}, \bibinfo {author} {\bibfnamefont {Q.}~\bibnamefont {Xiao}}, \bibinfo {author} {\bibfnamefont {S.}~\bibnamefont {Smith}}, \bibinfo {author} {\bibfnamefont {R.}~\bibnamefont {Abernathey}},\ and\ \bibinfo {author} {\bibfnamefont {A.~R.}\ \bibnamefont {Gray}},\ }\bibfield  {title} {\bibinfo {title} {Vertical fluxes conditioned on vorticity and strain reveal submesoscale ventilation},\ }\href@noop {} {\bibfield  {journal} {\bibinfo  {journal} {J. Phys. Oceanogr.}\ }\textbf {\bibinfo {volume} {51}},\ \bibinfo {pages} {2883} (\bibinfo {year} {2021})}\BibitemShut {NoStop}%
\bibitem [{\citenamefont {Bourgoin}\ \emph {et~al.}(2006)\citenamefont {Bourgoin}, \citenamefont {Ouellette}, \citenamefont {Xu}, \citenamefont {Berg},\ and\ \citenamefont {Bodenschatz}}]{bourgoin_etal_2006}%
  \BibitemOpen
  \bibfield  {author} {\bibinfo {author} {\bibfnamefont {M.}~\bibnamefont {Bourgoin}}, \bibinfo {author} {\bibfnamefont {N.~T.}\ \bibnamefont {Ouellette}}, \bibinfo {author} {\bibfnamefont {H.}~\bibnamefont {Xu}}, \bibinfo {author} {\bibfnamefont {J.}~\bibnamefont {Berg}},\ and\ \bibinfo {author} {\bibfnamefont {E.}~\bibnamefont {Bodenschatz}},\ }\bibfield  {title} {\bibinfo {title} {The role of pair dispersion in turbulent flow},\ }\href@noop {} {\bibfield  {journal} {\bibinfo  {journal} {Science}\ }\textbf {\bibinfo {volume} {311}},\ \bibinfo {pages} {835} (\bibinfo {year} {2006})}\BibitemShut {NoStop}%
\bibitem [{\citenamefont {LaCasce}(2008)}]{LaCasce2008}%
  \BibitemOpen
  \bibfield  {author} {\bibinfo {author} {\bibfnamefont {J.~H.}\ \bibnamefont {LaCasce}},\ }\bibfield  {title} {\bibinfo {title} {Statistics from {Lagrangian} observations},\ }\href@noop {} {\bibfield  {journal} {\bibinfo  {journal} {Prog. Oceanogr.}\ }\textbf {\bibinfo {volume} {77}},\ \bibinfo {pages} {1} (\bibinfo {year} {2008})}\BibitemShut {NoStop}%
\bibitem [{\citenamefont {Cencini}\ and\ \citenamefont {Vulpiani}(2013)}]{CV2013}%
  \BibitemOpen
  \bibfield  {author} {\bibinfo {author} {\bibfnamefont {M.}~\bibnamefont {Cencini}}\ and\ \bibinfo {author} {\bibfnamefont {A.}~\bibnamefont {Vulpiani}},\ }\bibfield  {title} {\bibinfo {title} {Finite size {Lyapunov} exponent: review on applications},\ }\href@noop {} {\bibfield  {journal} {\bibinfo  {journal} {J. Phys. A: Math. Theor.}\ }\textbf {\bibinfo {volume} {46}},\ \bibinfo {pages} {254019} (\bibinfo {year} {2013})}\BibitemShut {NoStop}%
\bibitem [{\citenamefont {Artale}\ \emph {et~al.}(1997)\citenamefont {Artale}, \citenamefont {Boffetta}, \citenamefont {Celani}, \citenamefont {Cencini},\ and\ \citenamefont {Vulpiani}}]{ABCCV1997}%
  \BibitemOpen
  \bibfield  {author} {\bibinfo {author} {\bibfnamefont {V.}~\bibnamefont {Artale}}, \bibinfo {author} {\bibfnamefont {G.}~\bibnamefont {Boffetta}}, \bibinfo {author} {\bibfnamefont {A.}~\bibnamefont {Celani}}, \bibinfo {author} {\bibfnamefont {M.}~\bibnamefont {Cencini}},\ and\ \bibinfo {author} {\bibfnamefont {A.}~\bibnamefont {Vulpiani}},\ }\bibfield  {title} {\bibinfo {title} {Dispersion of passive tracers in closed basins: beyond the diffusion coefficient},\ }\href@noop {} {\bibfield  {journal} {\bibinfo  {journal} {Phys. Fluids A}\ }\textbf {\bibinfo {volume} {9}},\ \bibinfo {pages} {3162} (\bibinfo {year} {1997})}\BibitemShut {NoStop}%
\bibitem [{\citenamefont {Boffetta}\ \emph {et~al.}(2000)\citenamefont {Boffetta}, \citenamefont {Celani}, \citenamefont {Lacorata},\ and\ \citenamefont {Vulpiani}}]{boffettaetal2000}%
  \BibitemOpen
  \bibfield  {author} {\bibinfo {author} {\bibfnamefont {G.}~\bibnamefont {Boffetta}}, \bibinfo {author} {\bibfnamefont {A.}~\bibnamefont {Celani}}, \bibinfo {author} {\bibfnamefont {G.}~\bibnamefont {Lacorata}},\ and\ \bibinfo {author} {\bibfnamefont {A.}~\bibnamefont {Vulpiani}},\ }\bibfield  {title} {\bibinfo {title} {The predictability problem in systems with an uncertainty in the evolution law},\ }\href@noop {} {\bibfield  {journal} {\bibinfo  {journal} {J. Phys. A: Math. Gen.}\ }\textbf {\bibinfo {volume} {33}},\ \bibinfo {pages} {1313} (\bibinfo {year} {2000})}\BibitemShut {NoStop}%
\bibitem [{\citenamefont {Lacorata}\ \emph {et~al.}(2019)\citenamefont {Lacorata}, \citenamefont {Corrado}, \citenamefont {Falcini},\ and\ \citenamefont {Santoleri}}]{lacorata2019}%
  \BibitemOpen
  \bibfield  {author} {\bibinfo {author} {\bibfnamefont {G.}~\bibnamefont {Lacorata}}, \bibinfo {author} {\bibfnamefont {R.}~\bibnamefont {Corrado}}, \bibinfo {author} {\bibfnamefont {F.}~\bibnamefont {Falcini}},\ and\ \bibinfo {author} {\bibfnamefont {R.}~\bibnamefont {Santoleri}},\ }\bibfield  {title} {\bibinfo {title} {{FSLE} analysis and validation of lagrangian simulations based on satellite-derived globcurrent velocity data},\ }\href@noop {} {\bibfield  {journal} {\bibinfo  {journal} {Remote Sens. Environ.}\ }\textbf {\bibinfo {volume} {221}},\ \bibinfo {pages} {136} (\bibinfo {year} {2019})}\BibitemShut {NoStop}%
\bibitem [{\citenamefont {Lapeyre}(2002)}]{lapeyre2002}%
  \BibitemOpen
  \bibfield  {author} {\bibinfo {author} {\bibfnamefont {G.}~\bibnamefont {Lapeyre}},\ }\bibfield  {title} {\bibinfo {title} {Characterization of finite-time {Lyapunov} exponents and vectors in two-dimensional turbulence},\ }\href@noop {} {\bibfield  {journal} {\bibinfo  {journal} {Chaos}\ }\textbf {\bibinfo {volume} {12}},\ \bibinfo {pages} {688} (\bibinfo {year} {2002})}\BibitemShut {NoStop}%
\bibitem [{\citenamefont {Cencini}\ \emph {et~al.}(2010)\citenamefont {Cencini}, \citenamefont {Cecconi},\ and\ \citenamefont {Vulpiani}}]{CCV2010}%
  \BibitemOpen
  \bibfield  {author} {\bibinfo {author} {\bibfnamefont {M.}~\bibnamefont {Cencini}}, \bibinfo {author} {\bibfnamefont {F.}~\bibnamefont {Cecconi}},\ and\ \bibinfo {author} {\bibfnamefont {A.}~\bibnamefont {Vulpiani}},\ }\href@noop {} {\emph {\bibinfo {title} {Chaos: from simple models to complex systems}}}\ (\bibinfo  {publisher} {World Scientific, Singapore},\ \bibinfo {year} {2010})\BibitemShut {NoStop}%
\bibitem [{\citenamefont {Benettin}\ \emph {et~al.}(1980)\citenamefont {Benettin}, \citenamefont {Galgani}, \citenamefont {Giorgilli},\ and\ \citenamefont {Strelcyn}}]{benettinetal1980}%
  \BibitemOpen
  \bibfield  {author} {\bibinfo {author} {\bibfnamefont {G.}~\bibnamefont {Benettin}}, \bibinfo {author} {\bibfnamefont {L.}~\bibnamefont {Galgani}}, \bibinfo {author} {\bibfnamefont {A.}~\bibnamefont {Giorgilli}},\ and\ \bibinfo {author} {\bibfnamefont {J.-M.}\ \bibnamefont {Strelcyn}},\ }\bibfield  {title} {\bibinfo {title} {{Lyapunov} characteristic exponents for smooth dynamical systems and for {Hamiltonian} systems; a method for computing all of them. {Part} 1: Theory},\ }\href@noop {} {\bibfield  {journal} {\bibinfo  {journal} {Meccanica}\ }\textbf {\bibinfo {volume} {15}},\ \bibinfo {pages} {9} (\bibinfo {year} {1980})}\BibitemShut {NoStop}%
\bibitem [{\citenamefont {Cressman}\ \emph {et~al.}(2004)\citenamefont {Cressman}, \citenamefont {Davoudi}, \citenamefont {Goldburg},\ and\ \citenamefont {Schumacher}}]{cressmanetal2004}%
  \BibitemOpen
  \bibfield  {author} {\bibinfo {author} {\bibfnamefont {J.~R.}\ \bibnamefont {Cressman}}, \bibinfo {author} {\bibfnamefont {J.}~\bibnamefont {Davoudi}}, \bibinfo {author} {\bibfnamefont {W.~I.}\ \bibnamefont {Goldburg}},\ and\ \bibinfo {author} {\bibfnamefont {J.}~\bibnamefont {Schumacher}},\ }\bibfield  {title} {\bibinfo {title} {{Eulerian} and {Lagrangian studies} in surface flow turbulence},\ }\href@noop {} {\bibfield  {journal} {\bibinfo  {journal} {New J. Phys.}\ }\textbf {\bibinfo {volume} {6}},\ \bibinfo {pages} {53} (\bibinfo {year} {2004})}\BibitemShut {NoStop}%
\bibitem [{\citenamefont {Boffetta}\ \emph {et~al.}(2004)\citenamefont {Boffetta}, \citenamefont {Davoudi}, \citenamefont {B.~Eckhardt},\ and\ \citenamefont {Schumacher}}]{boffettaetal2004}%
  \BibitemOpen
  \bibfield  {author} {\bibinfo {author} {\bibfnamefont {G.}~\bibnamefont {Boffetta}}, \bibinfo {author} {\bibfnamefont {J.}~\bibnamefont {Davoudi}}, \bibinfo {author} {\bibfnamefont {B.}~\bibnamefont {B.~Eckhardt}},\ and\ \bibinfo {author} {\bibfnamefont {J.}~\bibnamefont {Schumacher}},\ }\bibfield  {title} {\bibinfo {title} {Lagrangian tracers on a surface flow: the role of time correlations},\ }\href@noop {} {\bibfield  {journal} {\bibinfo  {journal} {Phys. Rev. Lett.}\ }\textbf {\bibinfo {volume} {93}},\ \bibinfo {pages} {134501} (\bibinfo {year} {2004})}\BibitemShut {NoStop}%
\bibitem [{\citenamefont {Dhanagare}\ \emph {et~al.}(2014)\citenamefont {Dhanagare}, \citenamefont {Musacchio},\ and\ \citenamefont {Vincenzi}}]{dhanagareetal2014}%
  \BibitemOpen
  \bibfield  {author} {\bibinfo {author} {\bibfnamefont {A.}~\bibnamefont {Dhanagare}}, \bibinfo {author} {\bibfnamefont {S.}~\bibnamefont {Musacchio}},\ and\ \bibinfo {author} {\bibfnamefont {D.}~\bibnamefont {Vincenzi}},\ }\bibfield  {title} {\bibinfo {title} {Weak-strong clustering transition in renewing compressible flows},\ }\href@noop {} {\bibfield  {journal} {\bibinfo  {journal} {J. Fluid Mech.}\ }\textbf {\bibinfo {volume} {761}},\ \bibinfo {pages} {431} (\bibinfo {year} {2014})}\BibitemShut {NoStop}%
\bibitem [{\citenamefont {Pumir}(2017)}]{pumir2017}%
  \BibitemOpen
  \bibfield  {author} {\bibinfo {author} {\bibfnamefont {A.}~\bibnamefont {Pumir}},\ }\bibfield  {title} {\bibinfo {title} {Structure of the velocity gradient tensor in turbulent shear flows},\ }\href@noop {} {\bibfield  {journal} {\bibinfo  {journal} {Phys. Rev. Fluids}\ }\textbf {\bibinfo {volume} {2}},\ \bibinfo {pages} {074602} (\bibinfo {year} {2017})}\BibitemShut {NoStop}%
\bibitem [{\citenamefont {Yang}\ \emph {et~al.}(2022)\citenamefont {Yang}, \citenamefont {Fang}, \citenamefont {Fang}, \citenamefont {Pumir},\ and\ \citenamefont {Xu}}]{yangetal2022}%
  \BibitemOpen
  \bibfield  {author} {\bibinfo {author} {\bibfnamefont {P.-F.}\ \bibnamefont {Yang}}, \bibinfo {author} {\bibfnamefont {J.}~\bibnamefont {Fang}}, \bibinfo {author} {\bibfnamefont {L.}~\bibnamefont {Fang}}, \bibinfo {author} {\bibfnamefont {A.}~\bibnamefont {Pumir}},\ and\ \bibinfo {author} {\bibfnamefont {H.}~\bibnamefont {Xu}},\ }\bibfield  {title} {\bibinfo {title} {Low-order moments of the velocity gradient in homogeneous compressible turbulence},\ }\href@noop {} {\bibfield  {journal} {\bibinfo  {journal} {J. Fluid Mech.}\ }\textbf {\bibinfo {volume} {947}},\ \bibinfo {pages} {R1} (\bibinfo {year} {2022})}\BibitemShut {NoStop}%
\bibitem [{\citenamefont {Sinha}\ \emph {et~al.}(2019)\citenamefont {Sinha}, \citenamefont {Balwada}, \citenamefont {Tarshish},\ and\ \citenamefont {Abernathey}}]{sinha19}%
  \BibitemOpen
  \bibfield  {author} {\bibinfo {author} {\bibfnamefont {A.}~\bibnamefont {Sinha}}, \bibinfo {author} {\bibfnamefont {D.}~\bibnamefont {Balwada}}, \bibinfo {author} {\bibfnamefont {N.}~\bibnamefont {Tarshish}},\ and\ \bibinfo {author} {\bibfnamefont {R.}~\bibnamefont {Abernathey}},\ }\bibfield  {title} {\bibinfo {title} {Modulation of lateral transport by submesoscale flows and inertia-gravity waves},\ }\href@noop {} {\bibfield  {journal} {\bibinfo  {journal} {J. Adv. Mod. Earth Syst.}\ }\textbf {\bibinfo {volume} {11}},\ \bibinfo {pages} {1039} (\bibinfo {year} {2019})}\BibitemShut {NoStop}%
\bibitem [{\citenamefont {Dù}\ \emph {et~al.}(2024)\citenamefont {Dù}, \citenamefont {Smith},\ and\ \citenamefont {Bühler}}]{du2024}%
  \BibitemOpen
  \bibfield  {author} {\bibinfo {author} {\bibfnamefont {R.~S.}\ \bibnamefont {Dù}}, \bibinfo {author} {\bibfnamefont {K.~S.}\ \bibnamefont {Smith}},\ and\ \bibinfo {author} {\bibfnamefont {O.}~\bibnamefont {Bühler}},\ }\href@noop {} {\bibinfo {title} {Next-order balanced model captures submesoscale physics and statistics}} (\bibinfo {year} {2024}),\ \Eprint {https://arxiv.org/abs/2408.03422} {arXiv:2408.03422} \BibitemShut {NoStop}%
\end{thebibliography}%

\end{document}